\documentclass[letterpaper,12pt]{article}

\usepackage{pifont,fleqn}
\usepackage{amsmath}
\usepackage{graphicx,float,color}
\usepackage{hyperref}
\hypersetup{colorlinks = true, linkcolor = blue, urlcolor  = blue, citecolor = blue, anchorcolor = blue}
\usepackage{multirow}
\usepackage[numbers,sort&compress]{natbib}
\usepackage{braket}
\usepackage{amsfonts}
\usepackage{amsmath}
\usepackage{amssymb} 
\usepackage{graphicx}
\usepackage{epic}
\usepackage{eepic}
\usepackage{epsfig}
\usepackage{latexsym}
\usepackage{color}
\usepackage{url}  
\usepackage{caption}
\usepackage[caption=false]{subfig}
\usepackage{float}
\usepackage{bm}
\usepackage{geometry}
\usepackage{slashed}
\usepackage{lipsum}
\makeatletter
\def\ps@pprintTitle{%
 \let\@oddhead\@empty
 \let\@evenhead\@empty
 \def\@oddfoot{}%
 \let\@evenfoot\@oddfoot}
\makeatother
\usepackage{etoolbox}

\begin{document}

\begin{center}
\Large{\textbf{Clifford-based spectral action and renormalization group analysis of the gauge couplings}}  \\[1.0cm]
 
\large{Ufuk Aydemir}
\\[0.4cm]

\small{
\textit{School of Physics, Huazhong University of Science and Technology, Wuhan, Hubei 430074, \\P. R. China}}
\vspace{.2cm}

\end{center}
\vspace{.2cm}

\vspace{0.6cm}
\begin{abstract}
The \textit{Spectral Action Principle} in noncommutative geometry derives the actions of the Standard Model and General Relativity (along with several other gravitational terms) by reconciling them in a geometric setting, and hence offers an explanation for their common origin. However, one of the requirements in the minimal formalism, unification of the gauge coupling constants, is not satisfied, since the basic construction does not introduce anything new that can change the renormalization group (RG) running of the Standard Model. On the other hand, it has been recently argued that incorporating structure of the Clifford algebra into the finite part of the spectral triple, the main object that encodes the complete information of a noncommutative space, gives rise to five additional scalar fields in the basic framework. We investigate whether these scalars can help to achieve unification. We perform a RG analysis at the one-loop level, allowing possible mass values of these scalars to float from the electroweak scale to the putative unification scale. We show that out of twenty  configurations of mass hierarchy in total, there does \textit{not} exist even a single case that can lead to unification. In consequence, we confirm that the spectral action formalism requires a model-construction scheme beyond the (modified) minimal framework.
unification, Morita equivalence, spectral triple, Gelfand duality, renormalization group
\end{abstract}
   

\newpage

\section{Introduction\label{sec:intro}}

Noncommutative geometry~(NCG)\cite{Connes:1994yd,Connes:1995tu,Connes:1996gi,Connes:1990qp} reformulates concepts of geometry in terms of operator algebras, in similarity to the commutative case in which this reformulation is enabled by Gelfand duality~\cite{vanSuijlekom:2015iaa,GraciaBondia:2001tr}. Gelfand duality provides a one-to-one correspondence between locally compact Hausdorff topological spaces and commutative $C^{*}$ algebras. Therefore, in commutative spaces, the geometric properties of a manifold $M$ can be studied through, instead set of points, the algebraic properties of the commutative algebra of smooth functions $C^\infty(M)$ defined on it. This is generalized in NCG through an object referred to as \textit{spectral triple}, which encodes the complete information of a (noncommutative) space.   

A spectral triple ($\mathcal{A}, \mathcal{H}, \mathcal{D}$) is formed by an involutive algebra $\mathcal{A}$, a Hilbert space $\mathcal{H}$ on which the algebra acts as bounded operators, and a (possibly) unbounded self-adjoint operator $\mathcal{D}$ in $\mathcal{H}$, referred to as (generalized) Dirac operator.  ($\mathcal{A}, \mathcal{H}, \mathcal{D}$) determines a spectral geometry, based on the spectrum of the operator $\mathcal{D}$. From this set of data on the spectral triple, the original manifold as a metric space can be recovered; while information on the manifold is retrieved by the algebra, the metric is recovered by the Dirac operator. Spacetime in this picture is extended to a product of a continuous four dimensional manifold by a finite discrete space with noncommutative geometry.
                         
In the spectral action formalism in the NCG framework~\cite{Chamseddine:1996rw,Chamseddine:1996zu,Chamseddine:2006ep,Chamseddine:2007hz,Chamseddine:2010ud,Chamseddine:2012sw,Connes:2007book}\footnote{See Ref.~\cite{vanSuijlekom:2015iaa,Marcolli:2018uea,Lizzi:2018dah} for pedagogical reviews.}, for a given spectral triple ($\mathcal{A}, \mathcal{H}, \mathcal{D}$), the action is constructed in terms of the Dirac operator $\mathcal{D}$, through the \textit{spectral action}, which depends on the spectrum of $ \mathcal{D}$. In the basic version of this construction, which yields mainly the SM and a modified gravity model, the algebra is chosen as
$\mathcal{A} = C^\infty(M)\otimes\mathcal{A}_F$ such that the finite part of the algebra is given as $\mathcal{A}_F=\mathbb{C}\oplus \mathbb{H}\oplus M_3(\mathbb{C})$, where $\mathbb{H}$ is the algebra of quaternions, and $M_3(\mathbb{C})$ is the algebra of $3\times 3$ matrices with elements in $\mathbb{C}$.  $\mathcal{H}$ in the spectral triple is the Hilbert space of spinors. Then, the spectral action derives the SM action and the action of GR, the latter of which consists of the Einstein-Hilbert and the cosmological constant terms; additionally, it yields a non-minimal coupling term between the Higgs boson and the curvature, the Gauss-Bonnet term, and the Weyl (or the conformal gravity) term. The SM particle content and the gauge structure are described by this geometry, where the Higgs boson appears as the connection in the extra discrete dimension. The gauge transformations (SM) eventuate from the unitary inner automorphisms of the algebra $\mathcal{A}$ while diffeomorphisms (GR) are encoded in the outer automorphisms.

Besides the appealing features of the spectral action formalism, there are various issues within the minimal construction. The most important of these is the requirement of gauge coupling unification dictated by the spectral action. Obviously, unification of the gauge couplings, in general, is an appealing quality in a model. However, it cannot be achieved by the particle content of the SM in the usual renormalization group (RG) running; assuming that the Wilsonian approach to RG running is valid below the scale at which noncommutativity becomes apparent, the RG running in the minimal spectral action is the same as in the SM, as well as the particle content (modulo a possible singlet scalar~\cite{Chamseddine:2012sw}). Therefore, an extension to the minimal spectral action is required, which in turn would correspond to physics beyond the Standard Model. Indeed, a recently proposed extension to the basic formalism accommodates (three versions of) a Pati-Salam type model~\cite{Chamseddine:2013rta} with gauge coupling unification (in contrast to the canonical Pati-Salam models in the literature in which unification is not a requirement~\cite{Pati:1974yy,Mohapatra:1974gc,Mohapatra:1974hk,Senjanovic:1975rk}).\footnote{Unification in the Pati-Salam models based on NCG can be realized in a variety of ways, as displayed in Refs.~\cite{Chamseddine:2015ata,Aydemir:2015nfa,Aydemir:2016xtj,Aydemir:2018cbb}.}

Recently, Lizzi and Kurkov in Ref~\cite{Kurkov:2017wmx}, based on the analysis of Ref.~\cite{DAndrea:2014ics}, argue that incorporating Clifford structure into the finite part of the spectral triple of the basic construction gives rise to five extra scalars, with the SM quantum numbers\footnote{The hypercharge normalization adopted in this paper is given as $Q_{em}=I_3^L+Y$.} $\Omega(1,1-1)$, $S(\bar{3},2,-\frac{1}{6})$, and $\Delta_{u,d,L}(3,1,\frac{2}{3})$,  where the $\Delta$ fields, although they have the same quantum numbers, have different Yukawa couplings in the primary version of the fermionic action.  It is pointed out in Ref.~\cite{Kurkov:2017wmx} that the fields $S$, $\Delta_u$, $\Delta_d$, and $\Delta_L$ do \textit{not} enter into the final version of the fermionic action.\footnote{Therefore, these fields, despite of the fact that they carry right quantum numbers to be called leptoquarks, are in fact by definition \textit{not} leptoquarks in this formalism. Note that the quantum number assignment alone does not necessarily guarantee that the scalars couple to lepton-quark pairs as leptoquarks by definition should do. Thus, these scalars are not relevant to discussion of scalar leptoquarks in relation to the ongoing LHC searches, for instance in the context of reported $B$-decay anomalies~\cite{Aaij:2014ora,Aaij:2017vbb,Aaij:2017uff,Aaij:2015yra,Aaij:2017deq,Huschle:2015rga,Sato:2016svk,Hirose:2016wfn,Lees:2012xj,Lees:2013uzd}, unlike the case for the scalar leptoquarks in the Pati-Salam models from NCG~\cite{Chamseddine:2013rta} which can indeed be relevant to these anomalies~\cite{Aydemir:2018cbb}.}
On the other hand, since all of these scalars appear in the bosonic part, they contribute to the RG running of the gauge couplings, which is the focal point of this work.

In this paper, we address the question whether these scalars can help satisfy the unification condition in this \textit{modified} minimal formalism. We perform a 1-loop renormalization group analysis by adopting the most general approach in which the extra scalars are allowed to acquire random order of masses between the electroweak scale and the presumed unification scale, \textit{i.e.}~the emergence scale of the spectral action.  We show that out of twenty possible configurations in total, depending on mass hierarchy of these additional scalars, there does \textit{not} exist even a single case that can lead to unification of the gauge coupling constants. 

The rest of the paper is organized as follows: In section~\ref{section2}, we review the minimal spectral action formalism in noncommutative geometry, whereas in section~\ref{section3} we briefly introduce the modified framework in which the extra scalars emerge. In section~\ref{section4}, we present the one-loop renormalization group analysis and display our results. Finally in~\ref{section5}, we end the paper with discussion and conclusions.

\section{The minimal spectral action}\label{section2}

 In this section, we briefly introduce the minimal spectral action formalism~\cite{Chamseddine:1996rw,Chamseddine:1996zu} that derives the SM action and the action of GR as well as various additional gravitational terms (hence providing a modified gravity model). Interested reader can consult to Ref.~\cite{Lizzi:2018dah} for a concise review, or to Refs.~\cite{vanSuijlekom:2015iaa,Marcolli:2018uea,Eckstein:2019dcb} for more comprehensive introductions.
 
Noncommutative geometry (NCG)~\cite{Connes:1994yd,Connes:1995tu,Connes:1996gi,Connes:1990qp} redefines concepts of geometry in operator algebraic terms, by replacing set of points in ordinary topology by (noncommutative) algebra of functions, in similarity to the commutative case where the link between geometry and algebra is provided by Gelfand duality~\cite{vanSuijlekom:2015iaa,GraciaBondia:2001tr}. Gelfand duality yields a one-to-one correspondence between locally compact Hausdorff topological spaces and commutative $C^{*}$ algebras, leading to interpretation of $C^{*}$ algebras as generalizations of topological spaces. Since also every unital, commutative $C^{*}$-algebra is isomorphic to $C^\infty(M)$, the algebra of continuous, complex-valued functions on a locally compact Hausdorff space $M$, the geometric information of a manifold $M$ can be recovered via the algebraic properties of the commutative algebra $C^\infty(M)$ defined on it. In the case of noncommutative geometry, generalized version of this correspondence is provided through spectral triple, which uniquely characterizes a (noncommutative) space.   
 
 The main element in NCG that encodes the complete information of a noncommutative space is the corresponding spectral triple, ($\mathcal{A}, \mathcal{H}, \mathcal{D}$), formed by an involutive algebra $\mathcal{A}$ of operators, a Hilbert space $\mathcal{H}$ (of fermionic states in our case), and a self-adjoint unbounded operator  $\mathcal{D}$, referred to as the (generalized) Dirac operator, with compact resolvent such that all commutators $[\mathcal{D},a]$ are bounded for $a\in\mathcal{A}$, inverse of which, $\mathcal{D}^{-1}$, is the analog of the infinitesimal unit of length $ds$ of ordinary geometry.
  
  
 In the spectral action formalism in the NCG framework~\cite{Chamseddine:1996rw,Chamseddine:1996zu,Chamseddine:2006ep,Chamseddine:2007hz,Chamseddine:2010ud,Chamseddine:2012sw,Connes:2007book}, the action is constructed in terms of the Dirac operator in the spectral triple, through ``the spectral action", the bosonic part of which depends on the spectrum of the Dirac operator $ \mathcal{D}$. In the basic version of the spectral action, which derives mainly the SM and GR actions, the algebra is chosen as 
\begin{equation}
\mathcal{A} = C^\infty(M)\otimes\mathcal{A}_F\;,
\end{equation}
 where $C^\infty(M)$ is the algebra of complex-valued differential functions on $M$, and $\mathcal{A}_F$ is the finite dimensional section,   
\begin{equation}
\mathcal{A}_F = \mathbb{C}\oplus \mathbb{H}\oplus M_3(\mathbb{C})\;.
\end{equation}
Here, $\mathbb{H}\subset M_2(\mathbb{C})$ is the algebra of quaternions, and $M_n(\mathbb{C})$ is the algebra of $n\times n$ matrices with elements in $\mathbb{C}$. The three terms in $\mathcal{A}$ lead to the group factors of the SM gauge symmetry, $U(1)$, $SU(2)$, and $SU(3)$, respectively. The main symmetry for the noncommutative space characterized by the spectral triple ($\mathcal{A}, \mathcal{H}, \mathcal{D}$) is the group $\mbox{Aut}(\mathcal{A})$ of the automorphisms of the algebra $\mathcal{A}$, which contains diffeomorphisms $\mbox{Diff}(M)$ of $M$ and the gauge symmetry transformations. The total gauge group eventuates from the inner automorphisms of the algebra $\mbox{Int}(\mathcal{A})\subset \mbox{Aut}(\mathcal{A})$, which is a subgroup of $\mbox{Aut}(\mathcal{A})$ of the form $\alpha(x)=bxb^{*}$ (for a unitarity $b \in \mathcal{A}$), whereas diffeomorphisms correspond to the quotient group; the outer automorphisms $\mbox{Out} (\mathcal{A})\equiv \mbox{Aut}(\mathcal{A})/\mbox{Int}(\mathcal{A})$. 
    
The product rule on the Hilbert space and the Dirac operator is given as      
\begin{equation}
\mathcal{H}= L^2(M,S)\otimes \mathcal{H}_F\;,\qquad \mathcal{D}=\slashed\partial_M\otimes 1_F+\gamma_5\otimes \mathcal{D}_F\;,
\end{equation}
where $(\mathcal{H}_F, \mathcal{D}_F)$ on $\mathcal{A}_F$ determines the finite section of the spectral geometry, whereas the continuous part, which is a spin Riemannian manifold, corresponds to square integrable spinors $L^2(M,S)$ and the Dirac operator $\slashed\partial_M$ of the Levi-Civita spin connection~$(w)$ on $M$,  which in terms of vierbein ($e$) is given as
\begin{equation}
\slashed\partial_M=\sqrt{-1} \gamma^{\mu}\nabla_{\mu}^s\;;\qquad \nabla_{\mu}^s=\partial_{\mu}+\dfrac{1}{4}w_{\mu}^{ab}(e) \gamma_{ab}\;,\qquad \gamma^{\mu}=\gamma^a e^{\mu}_a\;. 
\end{equation} 

The noncommutativity stems from the noncommutativity of the algebra, $\mathcal{A}_F$. For instance, if we turn off the finite part of the triple, $(\mathcal{A}_F, \mathcal{H}_F, \mathcal{D}_F)$, then we get the usual commutative case in which $(C^\infty(M),L^2(M,S), \slashed\partial_M)$ corresponds to the Riemannian compact spin manifold, as mentioned above. In this case, the group $\mbox{Diff}(M)$ is isomorphic to the group $\mbox{Aut}(C^\infty(M))$. Information on the ordinary metric can be recovered from this commutative case. Replacing the usual Riemannian manifold $M$ with the corresponding spectral triple does not cause any information loss. Points on $M$ are retrieved as the characters of the algebra $\mathcal{A}=C^\infty(M)$. The geodesic distance between points on $M$ is retrieved by the formula~\cite{Chamseddine:1996rw,Chamseddine:1996zu}
 \begin{equation}
 d(x,y)=\mbox{Sup} \{|a(x)-a(y)|\; ;\; a\in \mathcal{A}\;,\; \|[\mathcal{D},a]\|\leqslant1\}\;.
 \end{equation}
The spectral action is constructed in terms of the covariant Dirac operator in the spectral triple as~\cite{Chamseddine:1996rw,Chamseddine:1996zu}
\begin{equation}
\label{action}
\mathcal{S}=\mathcal{S}_F+\mathcal{S}_B= \left(J\psi, \mathcal{D}_A \psi\right)+\mbox{Tr}\left[\chi\left(\dfrac{\mathcal{D}_A}{\Lambda}\right)\right]\;,
\end{equation}
whose statement is referred to as the \textit{Spectral Action Principle}. The first and the second terms in Eq.~(\ref{action}) respectively denote the fermionic and bosonic actions. Tr denotes the trace in the Hilbert space $\mathcal{H}$.  $\chi$ is a cutoff function which serves as a regulator that selects the eigenvalues of covariant Dirac operator, $\mathcal{D}_A$, smaller than the cutoff $\Lambda$. $J$ is anti-unitary operator on $\mathcal{H}$,  called the real structure on the spectral triple, which can also be referred to as generalized charge conjugation, taking into account the presence of antiparticles. $\mathcal{D}_A$ is given by the following formula, which corresponds to taking the internal fluctuations of the metric,
\begin{equation}
\mathcal{D}_A=\mathcal{D}+A+JAJ^{\dag}\;,\quad A= \sum a_i \left[\mathcal{D},b_i\right]\;, \quad a_i\;,b_i \in \mathcal{A}\;,\quad A=A^*\;,
\end{equation}
where $\mathcal{D}$ is the unperturbed Dirac operator, and $A$ is a Hermitian one-form potential. The Dirac operator $\mathcal{D}$, as a differential operator of order one, satisfies so-called the \textit{first-order condition} (or the order-one condition)\footnote{However, see Refs.~\cite{Chamseddine:2013kza,Chamseddine:2013rta} for cases without the first-order condition.}

\begin{equation}
\left[\left[\mathcal{D},a\right], JbJ^{-1}\right]=0\;,\qquad \forall a,b\in \mathcal{A}\;.
\end{equation}
The derivative is defined as
\begin{equation}
da=\left[ \mathcal{D},a\right]\;,\qquad \forall a \in  \mathcal{A}\;. 
\end{equation}
Another important element in the formalism is ``grading" which is given as 
\begin{equation}
\Gamma= \gamma_5\otimes \gamma_F
\end{equation}
where $\gamma_5$ is the usual chirality operator for the continuous manifold, while $\gamma_F$ is for the finite part. The grading $\Gamma$ divides the Hilbert space $\mathcal{H}=sp(M)\otimes\mathcal{H}_F$ into ``left" and ``right", $\mathcal{H}=\mathcal{H}_L\oplus \mathcal{H}_R$, where $sp(M)$ corresponds to spinors in the continuous (spacetime) manifold. Because of this extra grading in the finite part, there is some over-counting of degrees of freedom, referred to as ``the fermion doubling problem" in the literature. One way of dealing with this issue is recently proposed in Ref.~\cite{DAndrea:2016hyl}, which utilizes Wick rotation in order  to get rid of spurious degrees of freedom. Note that in the spectral action formalism, $M$ is initially chosen to be a compact Riemannian manifold. Ref.~\cite{DAndrea:2016hyl} uses Wick rotation to introduce the Lorentzian signature into the theory, while at the same time resolving the doubling problem.  



In the fermionic action, it is necessary to construct the right form of the generalized Dirac operator, inserting the Yukawa couplings in the appropriate spots in $\mathcal{D}_F$,  in order to get the correct SM fermionic terms. Here, we only focus on the bosonic action as it is the relevant part to our discussion in this paper.   

The bosonic section of the spectral action, given in Eq.~(\ref{action}), can be put in a convenient form to exploit the Heat Kernel techniques~\cite{gilkey1994invariance,Vassilevich:2003xt} as
%
\begin{equation}
\label{bosonicaction}
\mathcal{S}_B=\mbox{Tr}\left[\chi\left(\dfrac{\mathcal{D}_A}{\Lambda}\right)\right]\simeq\mbox{Tr}\left[\chi\left(\dfrac{\mathcal{D}^2_A}{\Lambda^2}\right)\right]\;,
\end{equation}
which can be expanded in terms of moments of cut-off function $\chi$, $f_n$, in power series in terms of $\Lambda^{-1}$, as
\begin{equation}
\mathcal{S}_B\simeq\sum_n f_n \;a_n\left(\dfrac{\mathcal{D}_A^2}{\Lambda^2}\right)\;,
\end{equation}
where $a_n$ are the Seeley-de Witt coefficients which vanish for n odd. $f_n$ for n even are given as 
\begin{eqnarray}
f_0=\int^\infty_0  x\chi(x) dx\;,\quad f_2=\int^\infty_0  \chi(x) dx\;,\quad f_{2n+4}=(-1)^n \partial^n_x \chi(x) \bigg|_{x=0}\;,\quad \mbox{for } n\geqslant 0\;.
\end{eqnarray}
The final version of the bosonic spectral action is given as~\cite{Chamseddine:1996rw,Chamseddine:1996zu,Chamseddine:2006ep}
\begin{eqnarray}
\label{actionfinal}
\mathcal{S}_B &=&\int \bigg (\dfrac{1}{2\kappa_0^2}R+\alpha_{0\;} C_{\mu\nu\rho\sigma}C^{\mu\nu\rho\sigma}+\gamma_0+\tau_{0\;} R^{*}R^{*} \nonumber\\
&+& \dfrac{f_0}{2\pi^2}\left[ g_3^2\; G_{\mu\nu}^{i\;}G^{\mu\nu i}+g_2^2 F_{\mu\nu}^{m\;} F^{\mu\nu m}+\dfrac{5}{3} g_1^2 B_{\mu\nu} B^{\mu\nu }\right]\nonumber\\
&+& |D_{\mu} H|^2-\mu_0^2 |H|^2-\xi_0 R |H|^2+\lambda_0 |H|^4+O\left(\dfrac{1}{\Lambda^2}\right)\bigg)\sqrt{g}\;d^4x\;,
\end{eqnarray}
where $R$ is the Ricci scalar, $C_{\mu\nu\rho\sigma}$ is the Weyl tensor, $R^{*}R^{*}$ is the Gauss-Bonnet term, which is topological in four dimensions, and which integrates to the Euler characteristic. The constants $(\kappa_0,\alpha_0,\gamma_0,\tau_0,\mu_0,\xi_0,\lambda_0)$ in Eq.~(\ref{actionfinal}) are defined in terms of combinations of the original constants in the theory, whose exact definitions are not relevant to our discussion here.

For the canonical normalization of the gauge sector of the SM, as can be seen in the second line in Eq.~(\ref{actionfinal}), the following conditions are imposed.
\begin{equation}
\dfrac{g_3^2 f_0}{2\pi^2}=\dfrac{g_2^2 f_0}{2\pi^2}=\dfrac{5}{3}\dfrac{g_1^2 f_0}{2\pi^2}=\frac{1}{4}\;,
\end{equation}
which corresponds to the condition of unification of the gauge couplings 
\begin{equation}
g_3^2=g_2^2=\dfrac{5}{3} g_1^2\;,
\end{equation}
assumed  to be valid at a high energy scale, $M_U\sim \Lambda $.

Evidently, this condition cannot be satisfied in the minimal formalism in which the renormalization group running (below the scale at which NCG becomes apparent, $M_U$) is the same as in the SM with the same particle content, where unification is not realized. On the other hand, as we briefly discuss in the next section, incorporating the Clifford structure into the finite spectral triple introduces five new scalars into the picture~\cite{Kurkov:2017wmx}, which brings up the question whether these scalars can help satisfy the unification condition. However, the answer turns out to be negative, as displayed in this paper.

\newpage
\section{Modified minimal set-up and extra scalars}\label{section3}
In Ref.~\cite{DAndrea:2014ics}, D'Andrea and Dabrowski study generalizations of the notion of spin-manifold and Dirac spinors to noncommutative geometry by incorporating the Clifford structure into the finite spectral triple, $(\mathcal{A}_F, \mathcal{H}_F, \mathcal{D}_F)$, in the sense whether or not the finite spectral triple describes a (noncommutative) spin manifold, and the elements in the Hilbert space $\mathcal{H}$ can be characterized as ``spinors" in general manner (recall that the continuous part already knows about spin). They argue that in order for the necessary conditions for this generalization to be satisfied, \textit{i.e.} for the finite Hilbert space $\mathcal{H}_F$ of the spectral triple to be a Morita equivalence bimodule between the finite algebra $\mathcal{A}_F$ and the associated (complexified) Clifford algebra $\mathcal{C \ell}(M)$, additional terms should be included in the Dirac operator, as well as a modification to the standard grading in the minimal formalism. As a result of this procedure, new scalars emerge in the theory. 

Lizzi and Kurkov in~\cite{Kurkov:2017wmx}, based on the analysis of Ref.~\cite{DAndrea:2014ics}, investigate the extended scalar sector in this Clifford-based spectral action framework. They argue that in this modified scheme, in addition to the scalar sector of the minimal formalism which consists of the SM Higgs (and possibly a singlet  scalar~\cite{Chamseddine:2012sw}), there are five new scalars, three of which have the same quantum numbers but different couplings. The new scalars carry the following quantum numbers. 
\begin{equation}
S\left(\bar{3},2,-\dfrac{1}{6}\right)\;,\qquad \Delta_{u,d,L}\left(3,1,\dfrac{2}{3}\right)\;,\qquad \Omega\left(1,1,-1\right)\;,
\end{equation}
where the hypercharge normalization is $ Q_{em}=I_3^L+Y$. The subscripts $(u,d, L)$ indicate the type of fermion couplings (in the Dirac operator) that the corresponding scalar $\Delta_i$ possesses; couplings to the up-type and the down-type $SU(2)$ singlets, and $SU(2)$ doublets, respectively.

It is pointed out in Ref.~\cite{Kurkov:2017wmx} that the fields ($S$, $\Delta_u$, $\Delta_d$, $\Delta_L$) do \textit{not} enter into the final version of the fermionic action, although they appear in the bosonic part. Therefore, these fields, despite of the fact that they carry right quantum numbers to be called leptoquarks, are in fact by definition \textit{not} leptoquarks, in this formalism. Nevertheless, since they appear in the bosonic action, they are relevant to our discussion of RG running of the gauge couplings.

\section{Renormalization group analysis}\label{section4}
In this section, after we lay out preliminaries for the one-loop renormalization group (RG) running, we move on to the RG analysis with the aforementioned extra scalars. First, as a simple demonstration, we start with a special case  where the masses of the scalars are split between the electroweak scale and the unification scale, which is a common practice in the literature in the context of Grand Unified Theories (GUTs). Then, we investigate the most general case in which mass values of these five extra scalars are allowed to float between these two scales and  show that unification of the gauge couplings cannot be realized.

\subsection{Preliminaries}

For a given particle content, the gauge couplings corresponding to gauge group $G_i$ in an energy interval $\left[M_A,M_B\right]$ evolve under the one-loop RG running 
\begin{eqnarray}
\label{general}
\frac{1}{g_{i}^{2}(M_A)} - \dfrac{1}{g_{i}^2(M_B)}
\;=\; \dfrac{a_i}{8 \pi^2}\ln\dfrac{M_B}{M_A}
\;,
\end{eqnarray}
where the RG coefficients $a_i$ are given by \cite{Jones:1981we,Lindner:1996tf} 
\begin{eqnarray}
\label{1loopgeneral}
a_{i}
\;=\; -\frac{11}{3}C_{2}(G_i)
& + & \frac{2}{3}\sum_{R_f} T_i(R_f)\cdot d_1(R_f)\cdots d_n(R_f) \cr
& + & \frac{\eta}{3}\sum_{R_s} T_i(R_s)\cdot d_1(R_s)\cdots d_n(R_s)\;,
\end{eqnarray}
and the full gauge group is given as $G=G_i\otimes G_1\otimes...\otimes G_n$.

The summation in Eq.~(\ref{1loopgeneral}) is over irreducible chiral representations of fermions ($R_f$) and irreducible representations of scalars ($R_s$) in the second and the third terms, respectively. The coefficient $\eta$ is either 1 or 1/2, depending on whether the corresponding representation is complex or (pseudo) real, respectively. $d_j(R)$ is the dimension of the representation $R$ under the group $G_{j\neq i}$. $C_2(G_i)$ is the quadratic Casimir for the adjoint representation of the group $G_i$,
and $T_i$ is the Dynkin index of each representation. For $U(1)$ group, $C_2(G)=0$ and
\begin{equation}
\sum_{f,s}T \;=\; \sum_{f,s}Y^2\;,
\label{U1Dynkin}
\end{equation}
where $Y$ is the $U(1)$ charge.

The low energy data which we use as the boundary conditions in the RG running (in the $\overline{\mathrm{MS}}$ scheme) are
\cite{Patrignani:2016xqp,ALEPH:2005ab}
\begin{eqnarray}
\alpha^{-1}(M_Z) & = & 127.950\pm0.017\;,\cr
\alpha_s(M_Z) & = & 0.1182\pm0.0016\;,\cr
\sin^2\theta_W(M_Z) & = & 0.23129\pm0.00005\;,
\label{SMboundary}
\end{eqnarray}
at $M_Z=91.1876\pm 0.0021\,\mathrm{GeV}$, where we use only the central values throughout this work since the contribution from the deviations is negligible and does not change our results.

%
%

The obvious boundary/matching conditions to be imposed on the couplings at $M_U$ and $M_Z$ 
are:
\begin{eqnarray}
M_U & \;:\; & \sqrt{\frac{5}{3}}\;g_1(M_U) \;=\; g_2(M_U) \;=\; g_3(M_U) \;, \vphantom{\bigg|} 
\cr
M_Z & \;:\; & \frac{1}{e^2(M_Z)} \;=\; \frac{1}{g_1^2(M_Z)}+\frac{1}{g_2^2(M_Z)}\;.\vphantom{\Bigg|} 
\label{Matching0}
\end{eqnarray}
%

\begin{table}[!ht]
\vspace{1cm}
{\small
\begin{center}
\begin{tabular}{| l| l| l| l| l }
\hline
\# of fields & Configuration no. &Particle content & $\left(\Delta a_1,\Delta a_2, \Delta a_3\right) $  \\ 
\hline\hline
$ $ 
&$1$
& $S$ 
&  $\left(\dfrac{1}{18},\dfrac{1}{2},\dfrac{1}{3}\right) $ 
\vphantom{\Bigg|}\\
\cline{2-4}
1-p
&$2$
& $\Delta$ 
& $\left(\dfrac{4}{9},0,\dfrac{1}{6}\right) $ 
\vphantom{\Bigg|}\\
\cline{2-4}
&$3$
& $\Omega$ 
& $\left(\dfrac{1}{3},0,0\right) $ 
\vphantom{\Bigg|}\\
\hline
$ $ 
&$4$
& $S\Delta$ 
&  $\left(\dfrac{1}{2},\dfrac{1}{2},\dfrac{1}{2}\right) $  
\vphantom{\Bigg|}\\
\cline{2-4}
2-p
&$5$
& $S\Omega$ 
& $\left(\dfrac{7}{18},\dfrac{1}{2},\dfrac{1}{3}\right) $   
\vphantom{\Bigg|}\\
\cline{2-4}
&$6$
& $\Delta \Omega$ 
& $\left(\dfrac{7}{9},0,\dfrac{1}{6}\right) $   
\vphantom{\Bigg|}\\
\cline{2-4}
&$7$
& $\Delta \Delta$ 
& $\left(\dfrac{8}{9},0,\dfrac{1}{3}\right) $   
\vphantom{\Bigg|}\\
\hline
$ $ 
&$8$
& $\Delta\Delta\Delta$
& $\left(\dfrac{4}{3},0,\dfrac{1}{2}\right) $
\vphantom{\Bigg|}\\
\cline{2-4}
3-p
&$9$
& $\Delta\Delta S$ 
& $\left(\dfrac{17}{18},\dfrac{1}{2},\dfrac{2}{3}\right) $ 
\vphantom{\Bigg|}\\
\cline{2-4}
&$10$
& $\Delta\Delta\Omega$ 
& $\left(\dfrac{11}{9},0,\dfrac{1}{3}\right) $   
\vphantom{\Bigg|}\\
\cline{2-4}
&$11$
& $\Delta\Omega S$
& $\left(\dfrac{5}{6},\dfrac{1}{2},\dfrac{1}{2}\right) $   
\vphantom{\Bigg|}\\
\hline
$ $ 
&$12$
& $\Delta\Delta\Delta S$
&  $\left(\dfrac{25}{18},\dfrac{1}{2},\dfrac{5}{6}\right) $
\vphantom{\Bigg|}\\
\cline{2-4}
4-p
&$13$
& $\Delta\Delta\Delta\Omega$
& $\left(\dfrac{5}{3},0,\dfrac{1}{2}\right) $
\vphantom{\Bigg|}\\
\cline{2-4}
&$14$
& $\Delta\Delta\Omega S$
& $\left(\dfrac{23}{18},\dfrac{1}{2},\dfrac{2}{3}\right) $ 
\vphantom{\Bigg|}\\
\hline
5-p
&$15$
& $\Delta\Delta\Delta S \Omega$ 
& $\left(\dfrac{31}{18},\dfrac{1}{2},\dfrac{5}{6}\right) $  
\vphantom{\Bigg|}\\
\hline
\end{tabular}
\end{center}
}
\caption{\label{Configurations-special}
The field configurations and the corresponding modifications in the RG coefficients in the special case, in which we assume that the particles are split such that some of them acquire masses at the TeV-scale and the rest are heavy at the presumed unification scale.\\\\\
}
\end{table}
\begin{figure}
\vspace{-2.5cm}
\subfloat{\includegraphics[width=0.3\textwidth]{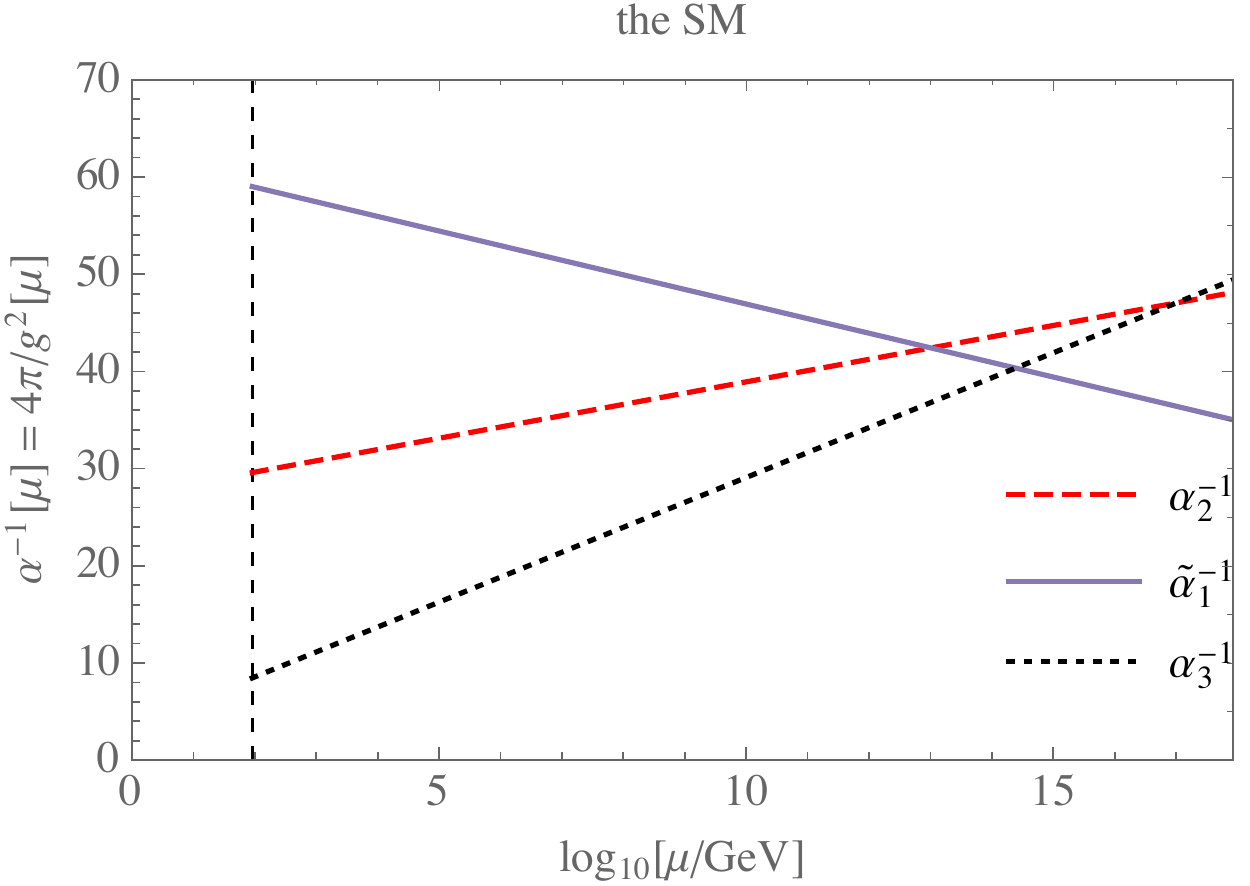} } \,
  \subfloat{\includegraphics[width=0.3\textwidth]{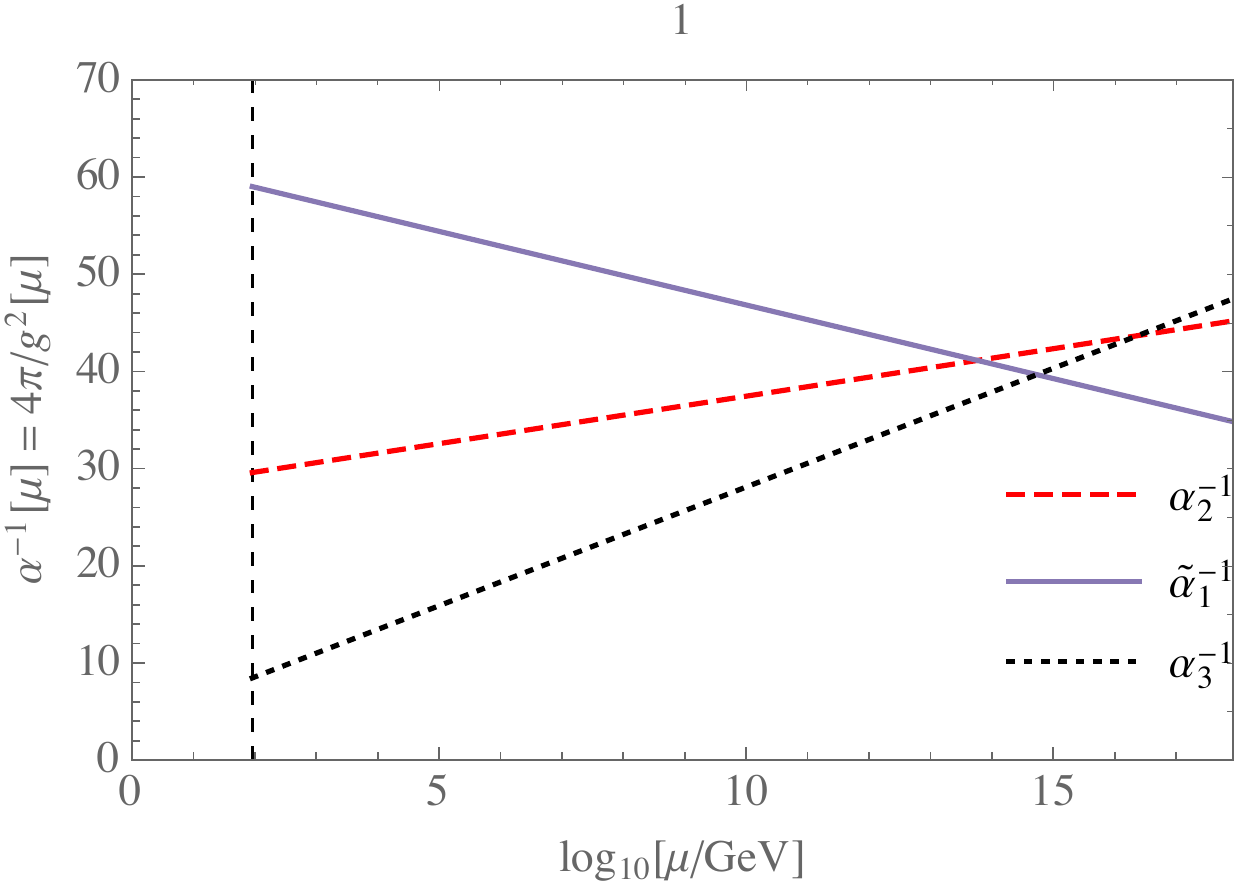} } \,
  \subfloat{\includegraphics[width=0.3\textwidth]{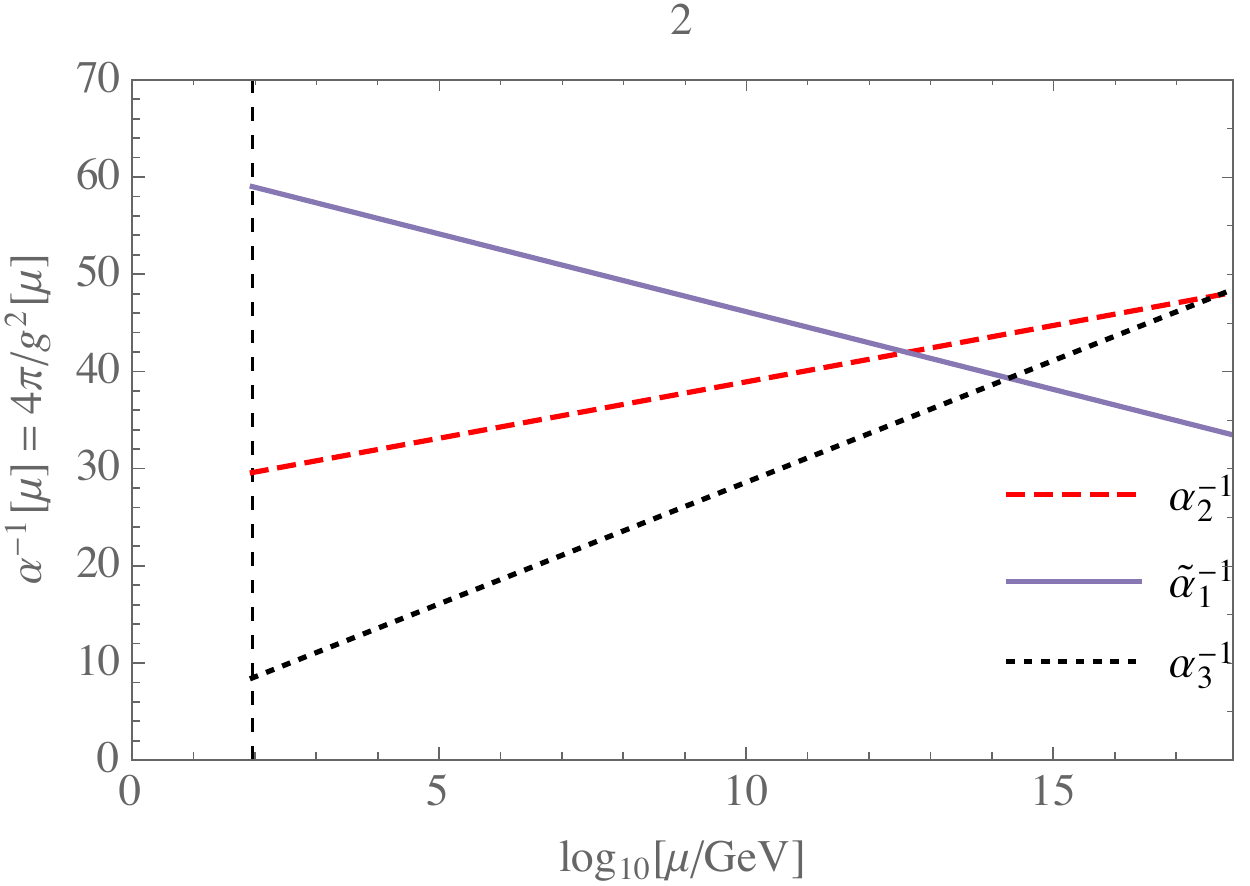} } \,
  \subfloat{\includegraphics[width=0.3\textwidth]{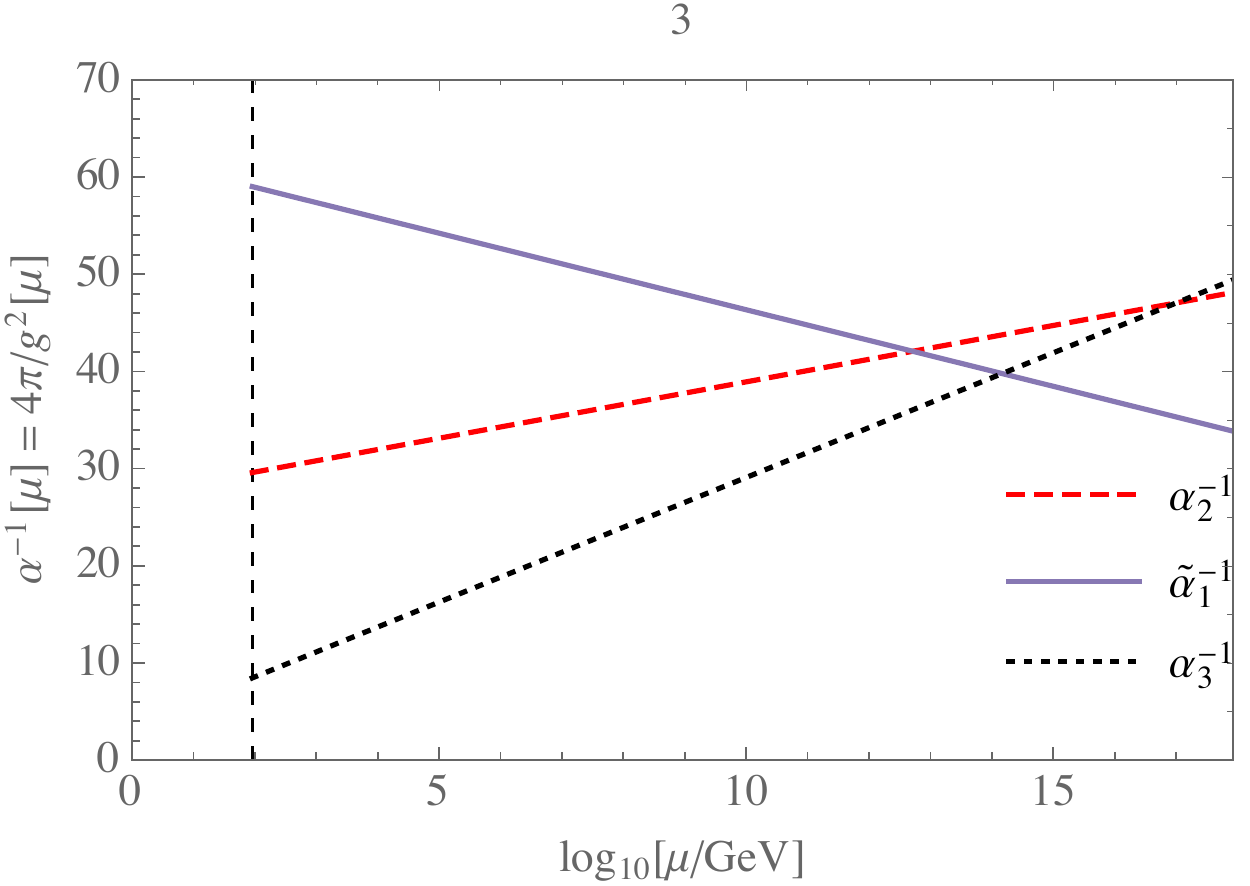} } \,
  \subfloat{\includegraphics[width=0.3\textwidth]{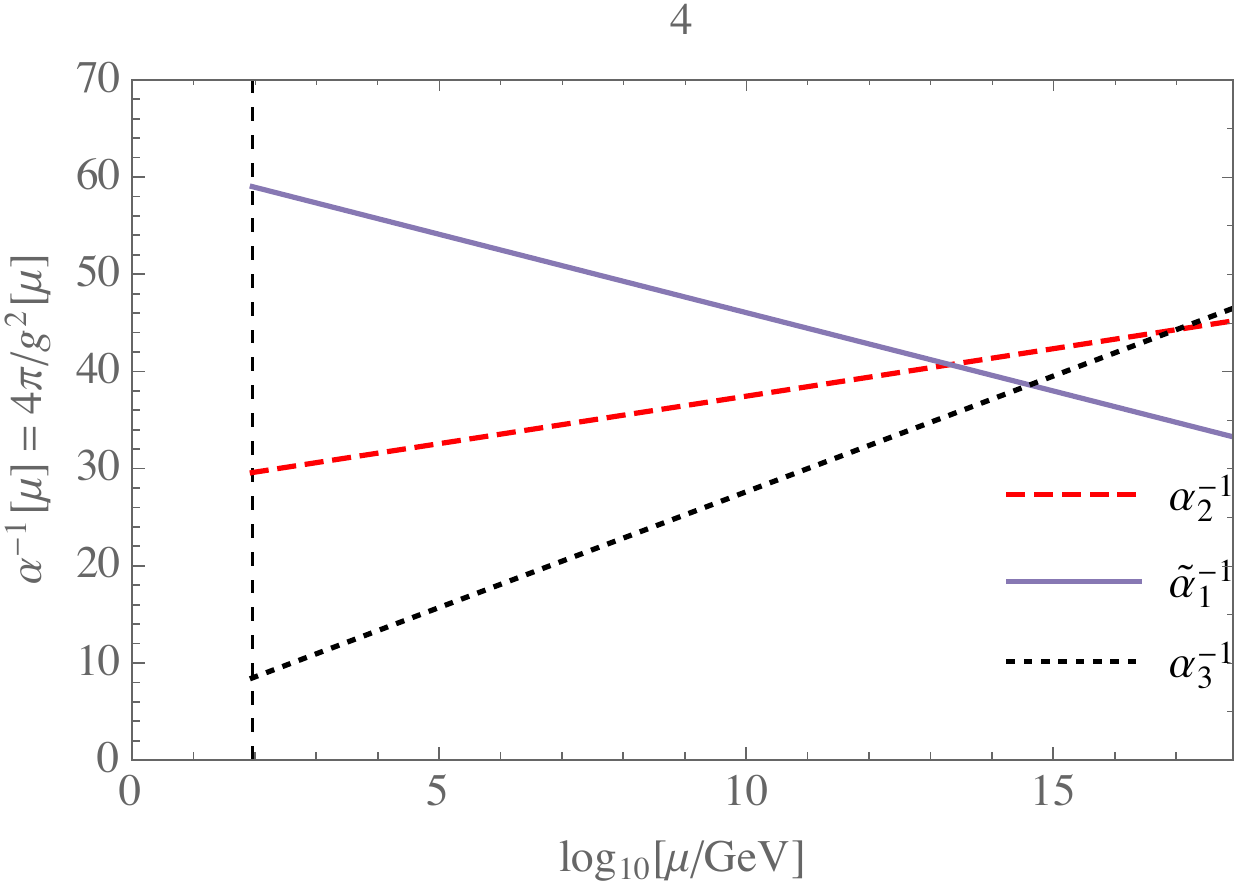} } \,
  \subfloat{\includegraphics[width=0.3\textwidth]{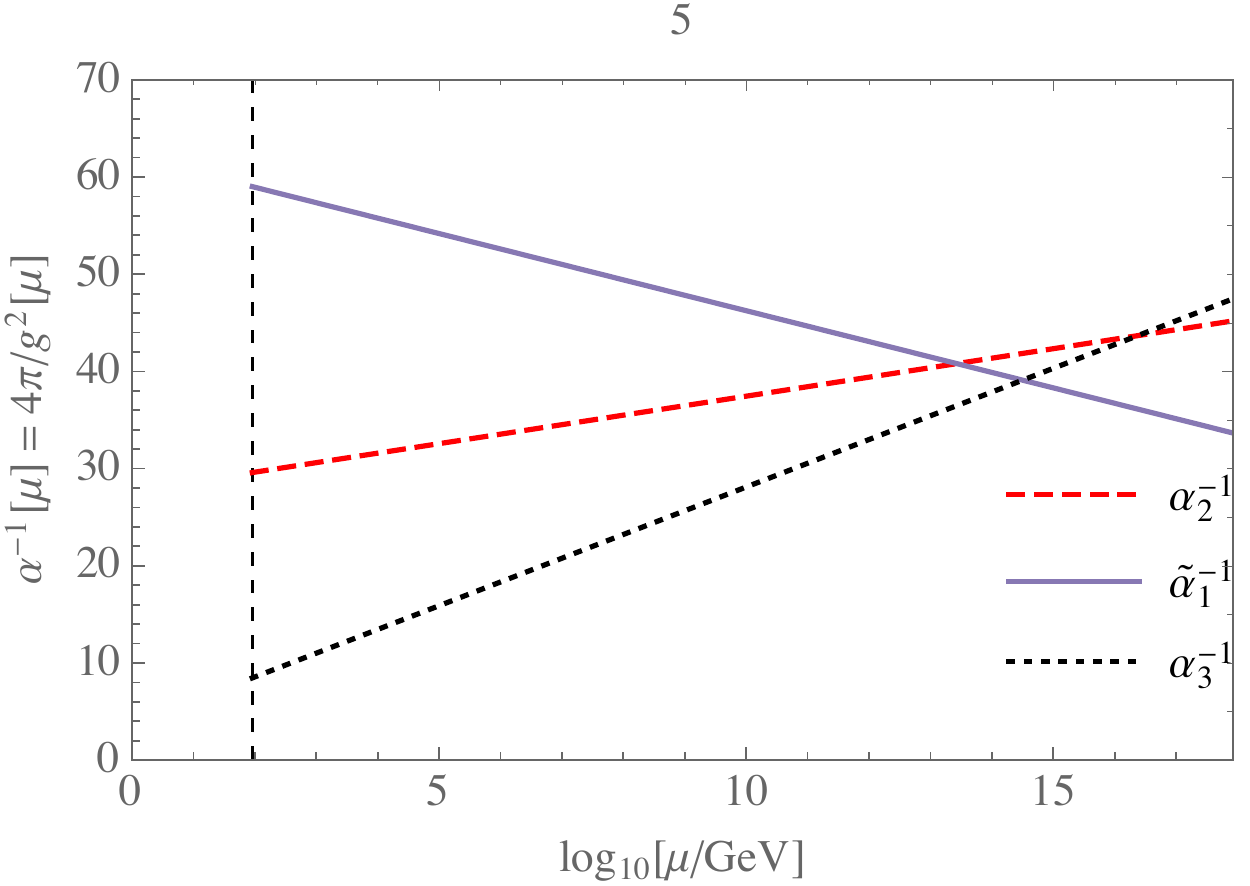} } \,
  \subfloat{\includegraphics[width=0.3\textwidth]{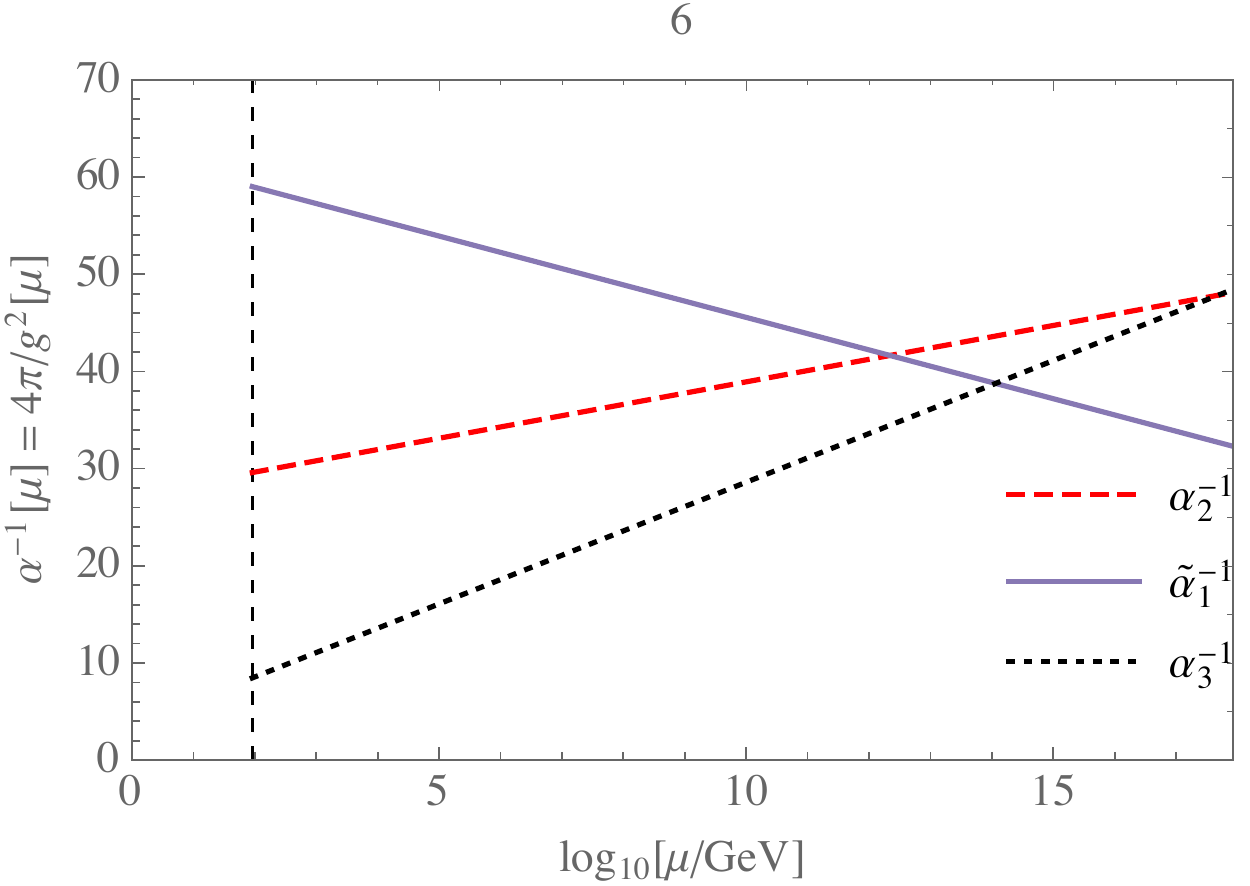} } \,
  \subfloat{\includegraphics[width=0.3\textwidth]{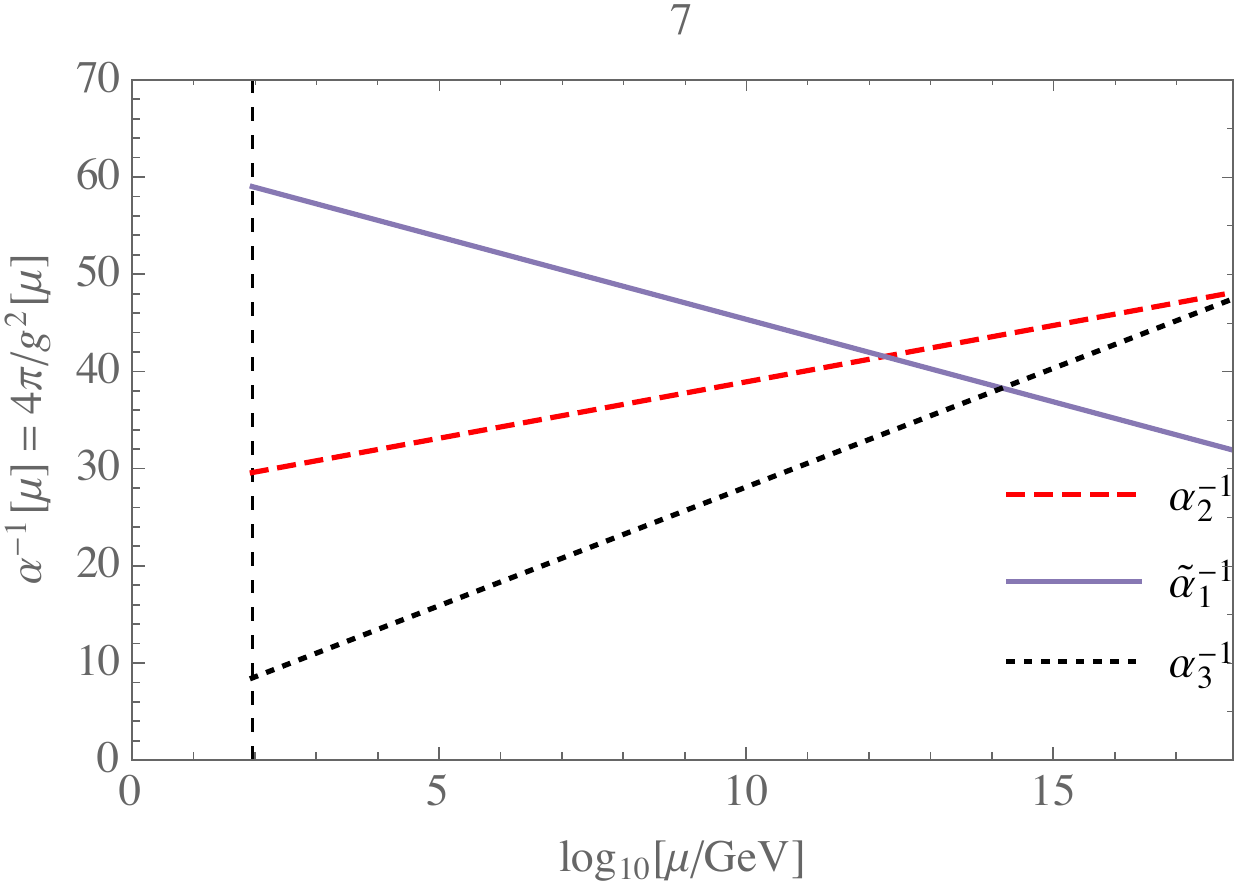} } \,
\subfloat{\includegraphics[width=0.3\textwidth]{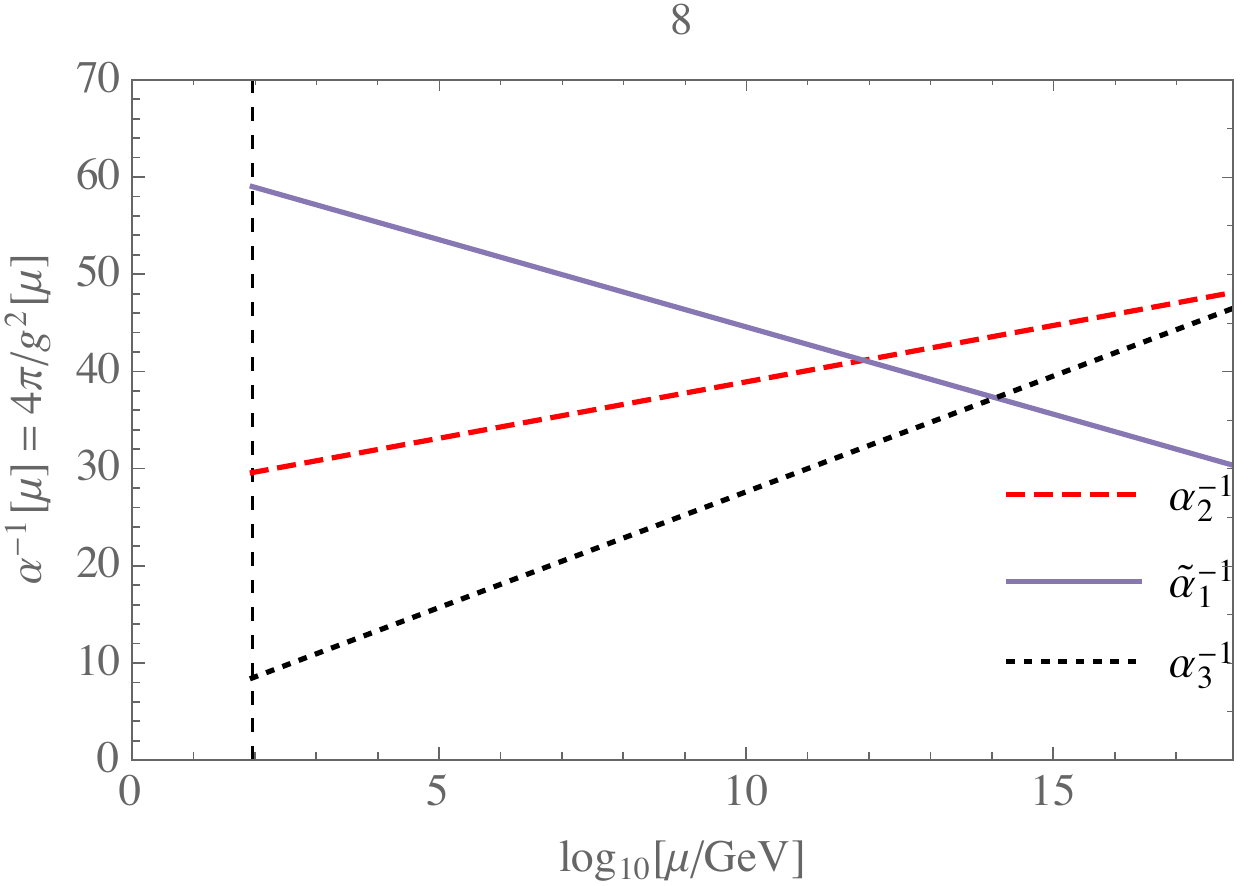} } \,
  \subfloat{\includegraphics[width=0.3\textwidth]{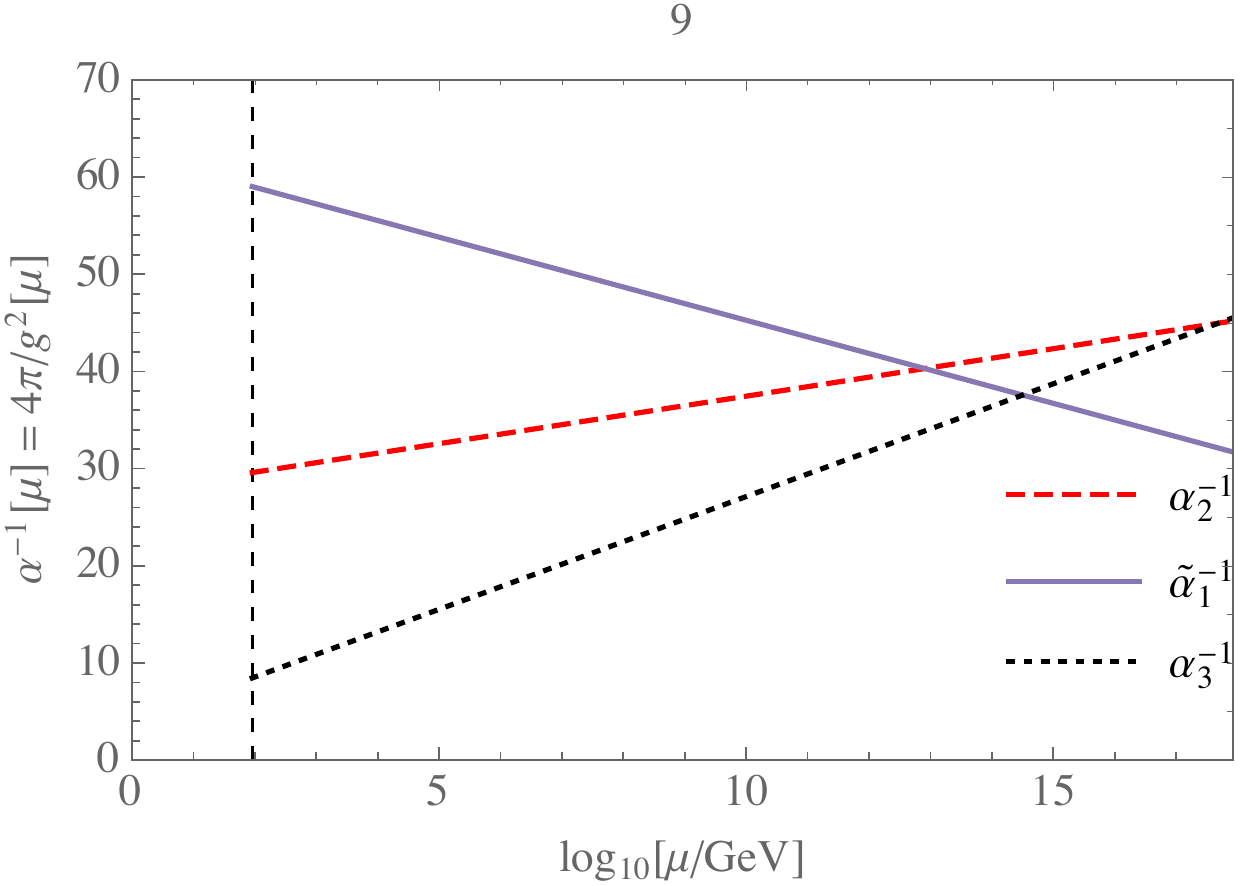} } \,
  \subfloat{\includegraphics[width=0.3\textwidth]{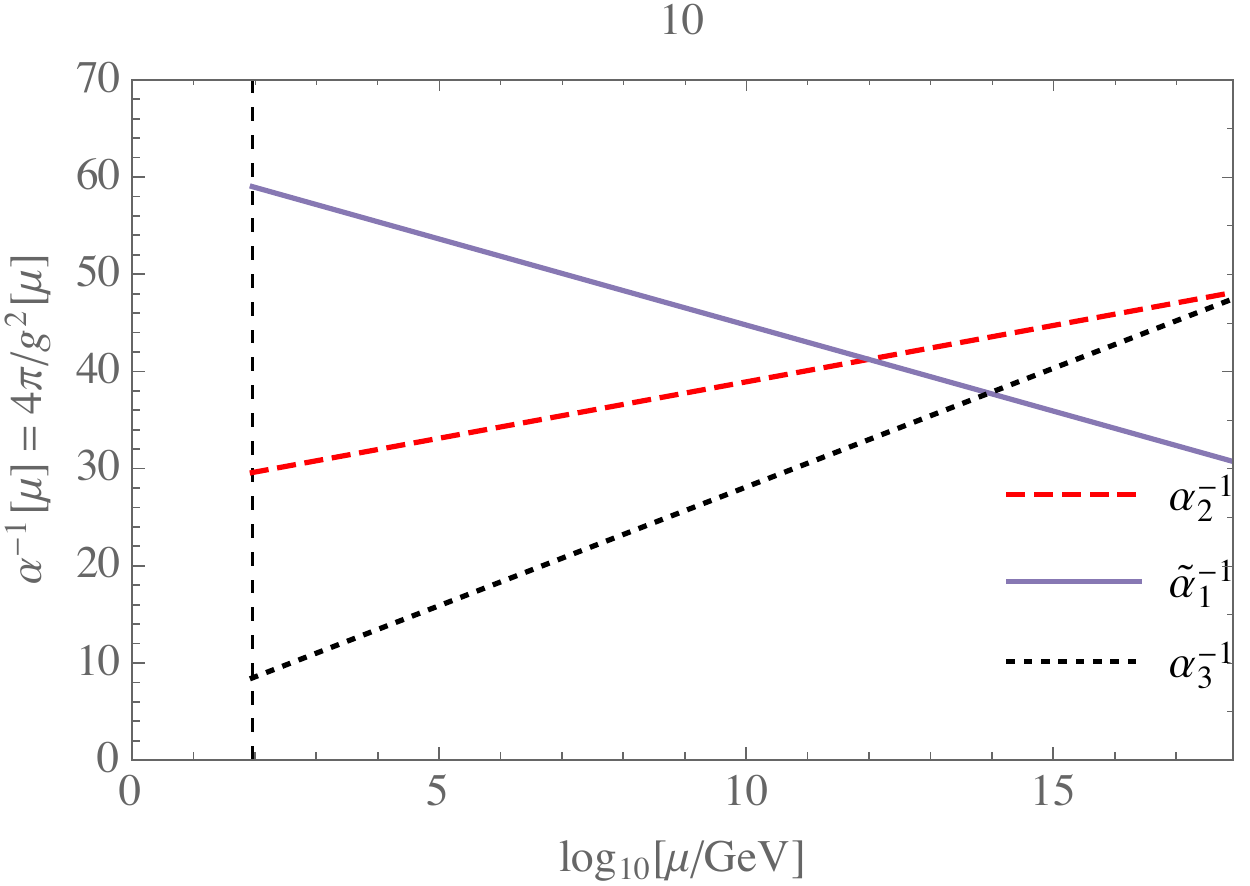} } \,
  \subfloat{\includegraphics[width=0.3\textwidth]{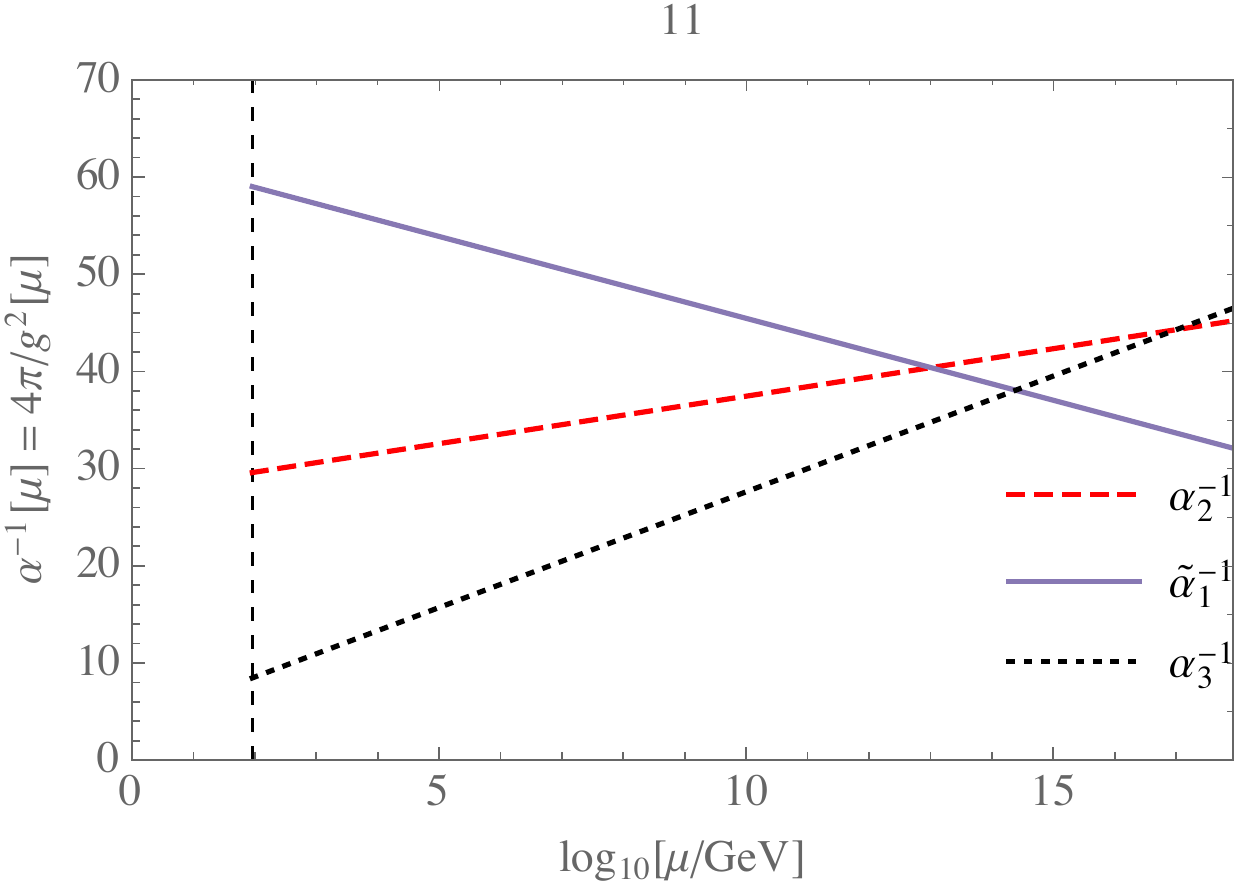} } \,
\subfloat{\includegraphics[width=0.3\textwidth]{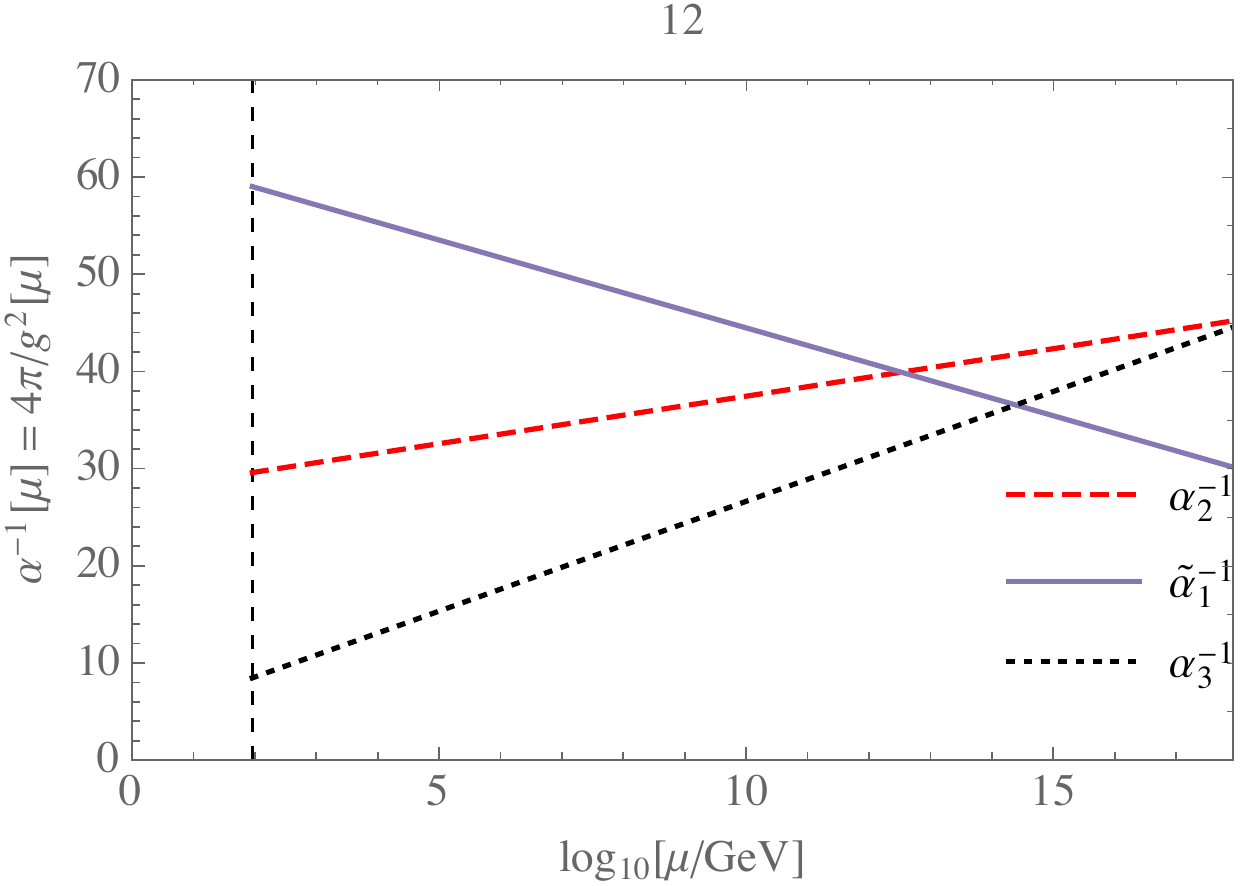} } \,
  \subfloat{\includegraphics[width=0.3\textwidth]{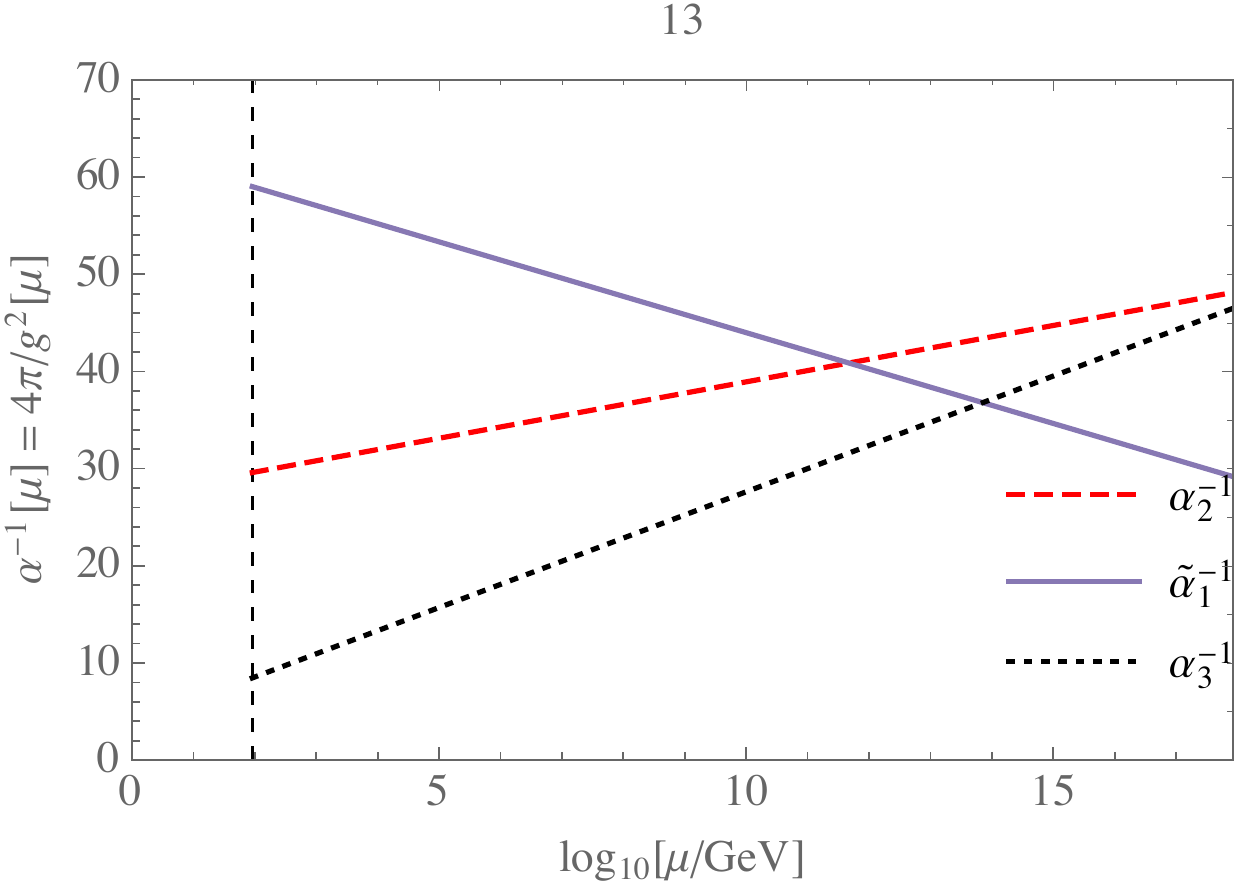} } \,
  \subfloat{\includegraphics[width=0.3\textwidth]{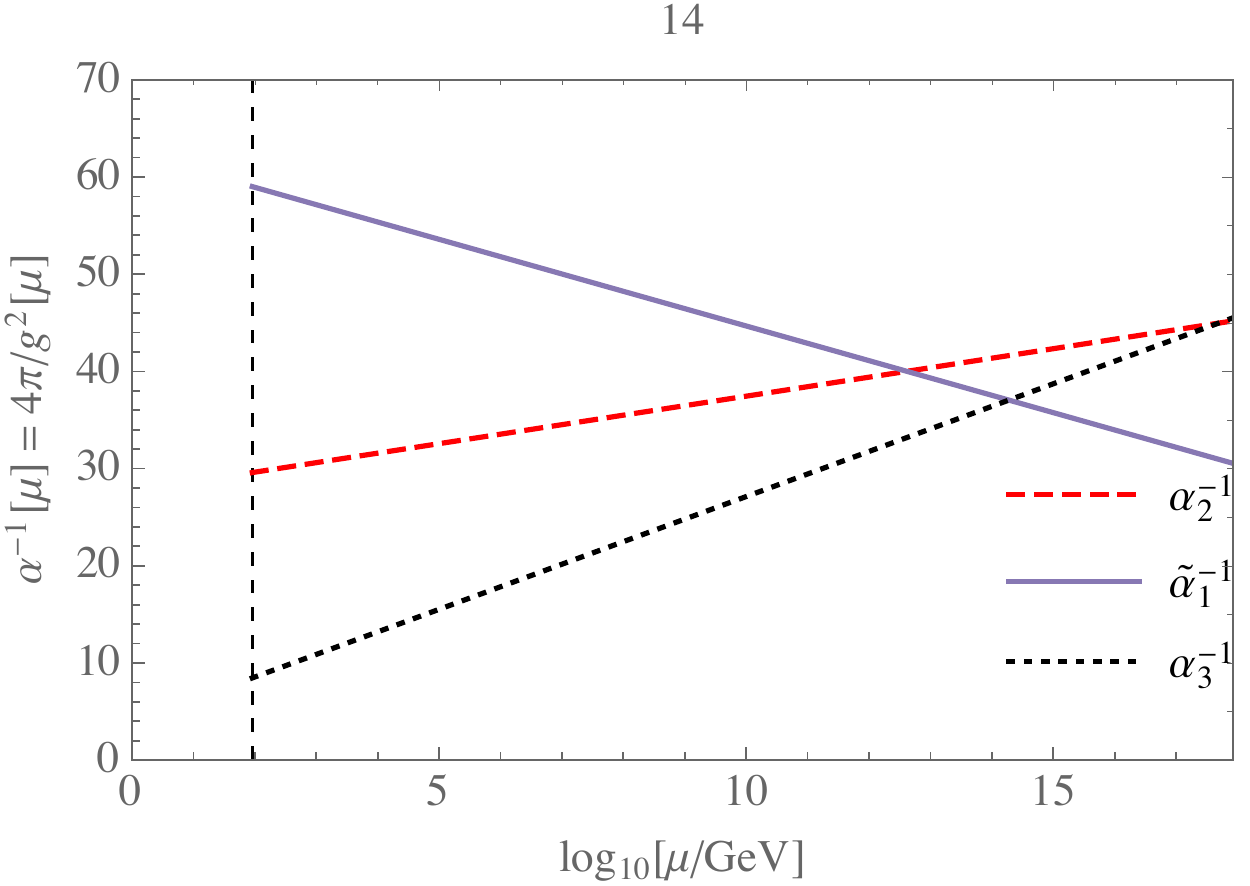} } \,
 \center \subfloat{\includegraphics[width=0.3\textwidth]{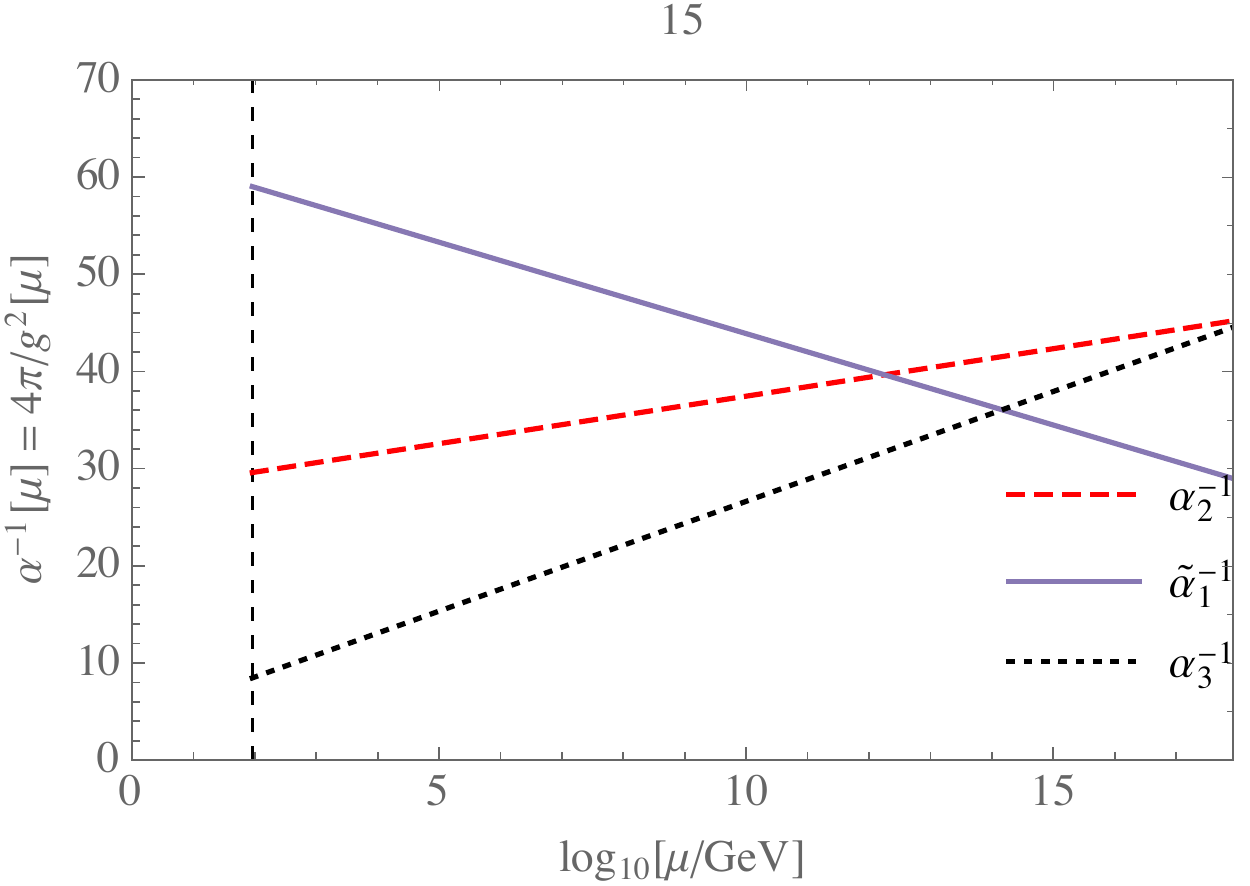} } \,
 \captionsetup{width=1.1\textwidth}
  \caption{Running of the gauge couplings for each scalar field configuration in the special case given in Table~\ref{Configurations-special}. The SM running is given in the first plot for comparison. The vertical dotted line correspond to the electroweak scale $M_Z$. For $\alpha_1^{-1}$, we plot the redefined quantity $\widetilde{\alpha}^{-1}_1\equiv \frac{3}{5}\alpha^{-1}_1$. Unification is realized in none of these cases.}
\label{RGrunning1}
\end{figure}

\subsection{Special case: TeV-scale scalars}

Before moving on to the general case where the masses of the scalars acquire values anywhere between the electroweak scale and the unification scale, we consider here a special case where the masses of the scalars are split between these two scales. The general case of course covers this specific version but we display it here anyway in part because it is convenient to use this case to illustrate to which extent these scalars modify the RG running of the SM, in part because TeV-scale scalars are relevant in terms of the LHC searches (see for instance Ref.~\cite{Aydemir:2018cbb,Aydemir:2016qqj}); thus, it is useful to inspect whether there is an improvement towards unification in each special case (given in Table~\ref{Configurations-special}), even though unification is not achieved.

Now, we will assume in each case that \textit{at least} one of these five scalars, 
\begin{equation}
S\left(\bar{3},2,-\dfrac{1}{6}\right)\;,\qquad \Delta_{u,d,L}\left(3,1,\dfrac{2}{3}\right)\;,\qquad \Omega\left(1,1,-1\right)\;,
\end{equation}
 are at the TeV-scale, and the rest of them are heavy at the unification scale, $M_U$. There are in total 15 possible configurations listed in Table~\ref{Configurations-special}.

The RG coefficients are given by
$a_i \;=\; a_i^\mathrm{SM}+\Delta a_i\;$,  $(i=1,2,3)$
where
\begin{eqnarray}
a_i^\mathrm{SM} & = & \left[\,\dfrac{41}{6}\,,\;-\dfrac{19}{6}\,,\;-7\,\right]\;,
\label{RGcoefI12}
\end{eqnarray}
and $\Delta a_i$ are the contributions from the new fields in each field configuration, \textit{i.e.} they are the relevant linear combinations of $\Delta a_i^{S}$, $\Delta a_i^{\Delta}$, and $\Delta a_i^{\Omega}$, which are, following Eq.~(\ref{1loopgeneral}), obtained as
\begin{eqnarray}
S&:&\qquad \Delta a_1^{S}=\frac{1}{18}\,,\qquad \Delta a_2^{S}=\frac{1}{2}\;,\qquad \Delta a_3^{S}=\frac{1}{3}\;,\nonumber\\
\Delta_{u,d,L}&:&\qquad  \Delta a_1^{\Delta}=\;\frac{4}{9}\;,\qquad  \Delta a_2^{\Delta}=0\;, \qquad  \Delta a_3^{\Delta}=\frac{1}{6}\;, \nonumber\\
\Omega &:&\qquad \Delta a_1^{\Omega}=\;\frac{1}{3}\;,\qquad \;\Delta a_2^{\Omega}=0\;,\qquad \Delta a_3^{\Omega}=0\;.
\end{eqnarray}
%


Out of 15 possible configurations listed in Table~\ref{Configurations-special}, unification is realized in none of them, as displayed in Fig.~\ref{RGrunning1}.
\subsection{The general case: Scalars with random order of masses}
In this section, we will now investigate the most general case in which these five extra scalars are allowed to acquire masses between the electroweak scale ($M_Z$) and the putative unification scale ($M_U$). Using Eq.~\eqref{1loopgeneral} with the low energy data and boundary conditions given in Eqs.~\eqref{SMboundary} and \eqref{Matching0}, the following equations are obtained.

\begin{eqnarray}
\label{equation1}
\lefteqn{
2\pi\left[\frac{3-8 \sin^2\theta_w (M_Z)}{\alpha(M_Z)}\right]} \vphantom{\Bigg|}\cr
& \quad = &
\mathcal{A}^\mathrm{VI}\ln\frac{M_5}{M_Z}+\mathcal{A}^{\mathrm{V}}\ln\frac{M_{4}}{M_5}
+\mathcal{A}^{\mathrm{IV}}\ln\frac{M_3}{M_{4}}
+\mathcal{A}^{\mathrm{III}}\ln\frac{M_2}{M_{3}}+\mathcal{A}^{\mathrm{II}}\ln\frac{M_1}{M_{2}}+\mathcal{A}^{\mathrm{I}}\ln\frac{M_U}{M_{1}}\;,\vphantom{\Bigg|}\cr
\lefteqn{2\pi\left[\frac{3}{\alpha(M_Z)}-\frac{8}{\alpha_s (M_Z)}\right]} \vphantom{\Bigg|}\cr
& \quad = & 
\mathcal{B}^\mathrm{VI}\ln\frac{M_5}{M_Z}+\mathcal{B}^{\mathrm{V}}\ln\frac{M_{4}}{M_5}
+\mathcal{B}^{\mathrm{IV}}\ln\frac{M_3}{M_{4}}
+\mathcal{B}^{\mathrm{III}}\ln\frac{M_2}{M_{3}}+\mathcal{B}^{\mathrm{II}}\ln\frac{M_1}{M_{2}}+\mathcal{B}^{\mathrm{I}}\ln\frac{M_U}{M_{1}}\;,\vphantom{\Bigg|}
\end{eqnarray}
where $M_i$ are the masses of five extra scalars, the Roman numerals label the corresponding energy intervals, and
\begin{equation}
\mathcal{A}\equiv 3 a_1-5 a_2\;, \qquad\qquad \mathcal {B}\equiv 3a_1+3a_2-8a_3\;.
\end{equation}
The RG coefficients and hence the values of the combinations above change accordingly each time a mass threshold is reached. 

There are in total 20 configurations of mass ordering, given in Table~ \ref{Configurations-general}. In none of these cases unification of couplings can be realized as we will show below. As can be seen from the last column in Table~ \ref{Configurations-general}, configurations $1^{\prime}-12^{\prime}$, $13^{\prime}-18^{\prime}$, and $19^{\prime}-20^{\prime}$ (where we use the superscript $(\prime)$ to indicate that the configurations belong to the general case) are equation-wise similar within themselves. Therefore, we will display one example for each group step by step, and since the necessary equations and results for each of 20 configurations are given in Table~ \ref{Configurations-general}, reader can easily reproduce the results in the other cases. Besides, we for comparison include in the last row in Table~\ref{Configurations-general} a hypothetical scenario with a scalar configuration different than ours that yields a positive result for unification.
\subsubsection{Case 1: Configuration $1^{\prime}$ ($\Delta \Delta \Delta S \Omega$)}
In this case, the mass ordering is given as 
\begin{equation}
M_U\geqslant M_{\Delta_u}\geqslant M_{\Delta_d}\geqslant M_{\Delta_l} \geqslant M_{S}\geqslant M_{\Omega} > M_{Z}\;,
\end{equation}
where the order of three fields $\Delta_{u,d,l}$ among themselves does not matter in the context of our discussion and hence any order combination among them is labeled as the same configuration in Table~\ref{Configurations-general}.  The intervals in Eq.~(\ref{equation1}) is labeled as
\begin{equation}
\left[M_U-M_1\right]\equiv \mathrm{I}\;,\quad \left[M_1-M_2\right]\equiv  \mathrm{II}\;,\;.\;.\;.\;,\;  \left[M_5-M_Z\right]\equiv  \mathrm{VI}\;,
\end{equation}
%
\begin{table}[!h]
{\footnotesize
\begin{center}
\begin{tabular}{| l| l| l|  l| l| }
\hline  
 Config. no. &Mass order  & Solutions & Contradictions &  \\ 
\hline\hline
$1^{\prime}$
& $\Delta\Delta\Delta S \Omega\;$ 
& $u=34+0.015 (e-d)$, $\quad a=183-b-c+1.3d -0.3e$    
&$u_{\mbox{{\tiny  max}}}=34,\;a_{\mbox{{\tiny  min}}}=62$
&$\times$
\vphantom{\Big|}
\\
$2^{\prime}$
& $\Delta\Delta\Delta \Omega S $  
& $u=34+0.015 (d-e)$, $\quad a=183-b-c+ 1.3e-0.3d $   
&$u_{\mbox{{\tiny  max}}}=35,\;a_{\mbox{{\tiny  min}}}=57$
&$\times$
\vphantom{\Big|}
\\
$3^{\prime}$
& $\Delta\Delta S \Delta \Omega$  
&$u=34+0.015 (e-c)$, $\quad a=183-b-d+ 1.3c-0.3e $   
&$u_{\mbox{{\tiny  max}}}=34,\;a_{\mbox{{\tiny  min}}}=92$
&$\times$
\vphantom{\Big|}
\\
$4^{\prime}$
& $\Delta\Delta S\Omega  \Delta $
&  $u=34+0.015 (d-c)$, $\quad a=183-b-e+ 1.3c-0.3d $  
&$u_{\mbox{{\tiny  max}}}=34,\;a_{\mbox{{\tiny  min}}}=92$
&$\times$
\vphantom{\Big|}
\\
$5^{\prime}$
&  $\Delta\Delta \Omega  \Delta S$ 
& $u=34+0.015 (c-e)$, $\quad a=183-b-d+ 1.3e-0.3c $    
&$u_{\mbox{{\tiny  max}}}=35,\;a_{\mbox{{\tiny  min}}}=57$
&$\times$
\vphantom{\Big|}
\\
$6^{\prime}$
&  $\Delta\Delta \Omega S  \Delta $ 
& $u=34+0.015 (c-d)$, $\quad a=183-b-e+ 1.3d-0.3c $  
&$u_{\mbox{{\tiny  max}}}=35,\;a_{\mbox{{\tiny  min}}}=80$
&$\times$
\vphantom{\Big|} 
\\
$7^{\prime}$
&  $\Delta S \Delta \Delta  \Omega $ 
& $u=34+0.015 (e-b)$, $\quad a=183-c-d+ 1.3b-0.3e $    
&$u_{\mbox{{\tiny  max}}}=34,\;a_{\mbox{{\tiny  min}}}=92$
&$\times$
\vphantom{\Big|}
\\
$8^{\prime}$
& $\Delta S \Delta \Omega  \Delta $ 
& $u=34+0.015 (d-b)$, $\quad a=183-c-e+ 1.3b-0.3d $ 
&$u_{\mbox{{\tiny  max}}}=34,\;a_{\mbox{{\tiny  min}}}=92$ 
&$\times$
\vphantom{\Big|}
\\
$9^{\prime}$
& $\Delta S \Omega \Delta  \Delta $
&$u=34+0.015 (c-b)$, $\quad a=183-d-e+ 1.3b-0.3c $ 
&$u_{\mbox{{\tiny  max}}}=34,\;a_{\mbox{{\tiny  min}}}=92$
&$\times$  
\vphantom{\Big|}
\\
$10^{\prime}$
&
 $\Delta \Omega \Delta  \Delta S $ 
&$u=34+0.015 (b-e)$, $\quad a=183-c-d+ 1.3e-0.3b $   
&$u_{\mbox{{\tiny  max}}}=35,\;a_{\mbox{{\tiny  min}}}=57$
&$\times$
\vphantom{\Big|}
\\
$11^{\prime}$
&
 $\Delta \Omega \Delta S \Delta  $ 
& $u=34+0.015 (b-d)$, $\quad a=183-c-e+ 1.3d-0.3b $    
&$u_{\mbox{{\tiny  max}}}=35,\;a_{\mbox{{\tiny  min}}}=80$
&$\times$
\vphantom{\Big|}
\\
$12^{\prime}$
& $\Delta \Omega S \Delta\Delta$
&  $u=34+0.015 (b-c)$, $\quad a=183-d-e+ 1.3c-0.3b $
&$u_{\mbox{{\tiny  max}}}=35,\;a_{\mbox{{\tiny  min}}}=92$
&$\times$
\vphantom{\Big|}
\\
$13^{\prime}$
& $ S \Delta\Delta\Delta\Omega$
&$u=34+0.015 (e-a)$, $\quad c=183-b-d+ 1.3a-0.3e $
&$u_{\mbox{{\tiny  max}}}=34,\;c_{\mbox{{\tiny  min}}}=92$
&$\times$
\vphantom{\Big|}
\\
$14^{\prime}$
&
 $ S \Delta\Delta\Omega \Delta$
&$u=34+0.015 (d-a)$, $\quad c=183-b-e+ 1.3a-0.3d $ 
&$u_{\mbox{{\tiny  max}}}=34,\;c_{\mbox{{\tiny  min}}}=92$
&$\times$
\vphantom{\Big|}
\\
$15^{\prime}$
&
 $ S \Omega \Delta\Delta\Delta $ 
& $u=34+0.015 (b-a)$, $\quad c=183-d-e+ 1.3a-0.3b $   
&$u_{\mbox{{\tiny  max}}}=34,\;c_{\mbox{{\tiny  min}}}=92$
&$\times$
\vphantom{\Big|}
\\
$16^{\prime}$
& $\Omega \Delta\Delta\Delta S $
& $u=34+0.015 (a-e)$, $ \quad c=183-b-d+1.3e -0.3a$ 
&$u_{\mbox{{\tiny  max}}}=35,\;c_{\mbox{{\tiny  min}}}=72$
&$\times$
\vphantom{\Big|}
\\
$17^{\prime}$
&  $\Omega \Delta\Delta S \Delta $
&  $u=34+0.015 (a-d)$, $\quad c=183-b-e+ 1.3d-0.3a $
&$u_{\mbox{{\tiny  max}}}=35,\;c_{\mbox{{\tiny  min}}}=139$
&$\times$
\vphantom{\Big|}
\\
$18^{\prime}$
& $\Omega S \Delta  \Delta  \Delta $
&$u=34+0.015 (a-b)$, $\quad c=183-d-e+ 1.3b-0.3a $   
&$u_{\mbox{{\tiny  max}}}=35,\;c_{\mbox{{\tiny  min}}}=102$
&$\times$
\vphantom{\Big|}
\\
$19^{\prime}$
& $ S \Delta\Omega \Delta\Delta $ 
& $u=34+0.015 (c-a)$, $\quad b=183-d-e+ 1.3a-0.3c $
&$u_{\mbox{{\tiny  max}}}=34,\;b_{\mbox{{\tiny  min}}}=92$
&$\times$
\vphantom{\Big|}
\\
$20^{\prime}$
&
$\Omega \Delta S \Delta  \Delta $
& $u=34+0.015 (a-c)$, $\quad b=183-d-e+ 1.3c-0.3a $   
&$u_{\mbox{{\tiny  max}}}=35,\;b_{\mbox{{\tiny  min}}}=102$
&$\times$
\vphantom{\Big|}
\\
\hline
\vphantom{\Big|}
Hypotheti- & $\Omega S S  S  \Delta $
&$u=35.4+0.011 (a-b)\;,\;$\; \textit{e.g.} $u=a=b=35.4\;,$ & $$ &$$ \\
cal positive &   &  $ c=-14.56-d-e\qquad\quad\quad\;\;\;\;$ c=9.8\;, d=e=5.5

 && $\checkmark$\\
 scenario &   &  $ \;\;\;\;\;\;\;+0.28a+0.72b \qquad\quad\quad\;\;$  && $$

\\
\hline
\end{tabular}
\end{center}
}
\caption{\label{Configurations-general}
The main result of this paper. The configurations of mass hierarchy and the corresponding equations in the general case, in which the extra scalars acquire random order of masses between the electroweak scale and the presumed unification scale, are given. The superscript $(\prime)$ indicates that the configurations belong to the general case. The mass orderings in the second column are given in the decreasing order from left to right. Similarly, $(u,a,b,c,d,e)$ denote mass scales as $u=\ln M_U/\mbox{GeV}$, $a = \ln M_1/\mbox{GeV}...$, where $M_U$ is the unification scale in each case, and $M_i$ is the mass of the scalar $i$, for $i=1,...5$. In the third column, we display the solutions of Eq.~(\ref{equation1}) for each case in a form convenient to observe the contradictions with the condition $u\geqslant a\geqslant b\geqslant c \geqslant d\geqslant e$. In \textit{none} of these 20 configurations, a positive result exits for gauge coupling unification, as detailed in the text. In the last row, we display for comparison a hypothetical scenario, in which there are ($\Omega, S, S, S, \Delta$) fields instead of our original ($\Omega, \Delta, \Delta, \Delta, S$), that yields gauge coupling unification for a range of values for $(a,b,c,d,e)$, for which an example-set of values is given.\\}
\end{table}
and $M_i$ denote the masses of the scalars  from the heaviest one to the lightest, as $i = 1,...5$.  We also introduce the following notation.
\begin{equation}
\label{notation}
u\equiv \ln\frac{M_U}{\mbox{GeV}}\;,\quad a\equiv \ln\frac{M_1}{\mbox{GeV}}\;,...\;,\quad e\equiv \ln\frac{M_5}{\mbox{GeV}}\;.
\end{equation}
%

\begin{table}[t]
\begin{center}
{\begin{tabular}{c|l|l}
\hline
$\vphantom{\Big|}$Interval & Active (extra) scalar dofs & RG coefficients  $\bigl[a_{1},a_{2},a_{3}\bigr]$
\\
\hline
$\vphantom{\Bigg|}$ I $(M_U-M_1)$
& $\Delta\Delta \Delta S \Omega$
& $\left[\dfrac{77}{9},-\dfrac{8}{3},-\dfrac{37}{6}\right]$
\\
\hline
$\vphantom{\Bigg|}$ II $(M_1-M_2)$
&$\Delta \Delta S \Omega$
& $\left[\dfrac{73}{9},-\dfrac{8}{3},-\dfrac{19}{3}\right]$
\\
\hline
$\vphantom{\Bigg|}$ III $(M_2-M_3)$
& $\Delta S \Omega$
& $\left[\dfrac{23}{3},-\dfrac{8}{3},-\dfrac{13}{2}\right]$
\\
\hline
$\vphantom{\Bigg|}$ IV $(M_3-M_4)$
& $ S \Omega$
& $\left[\dfrac{65}{9},-\dfrac{8}{3},-\dfrac{20}{3}\right]$
\\
\hline
$\vphantom{\Bigg|}$ V $(M_4-M_5)$
& $ \Omega$ 
& $\left[
\dfrac{43}{6},-\dfrac{19}{6},-7
\right]$
\\
\hline
$\vphantom{\Biggl|}$   VI $(M_5-M_Z)$
& $$
& $\left[
  \dfrac{41}{6},-\dfrac{19}{6},-7
  \right]$
\\
\hline
\end{tabular}}
\caption{\label{config1}The distribution of the new scalars among the energy intervals and the corresponding RG coefficients for configuration $1^{\prime}$ ($\Delta\Delta \Delta S \Omega$), \textit{i.e.} $M_U\geqslant M_{\Delta_u}\geqslant M_{\Delta_d}\geqslant M_{\Delta_l} \geqslant M_{S}\geqslant M_{\Omega} > M_{Z}$.}
\end{center}
\end{table}

Using Eq.~(\ref{equation1}) with the RG coefficients given in Table~\ref{config1} and definitions in Eq.~(\ref{notation}), after some fortunate cancellations, we obtain
\begin{eqnarray}
3265 &=& 117 u-4(a+b+c)+7d-3e\;,\nonumber\\
2289 &=& 67 u +d-e\;.
\end{eqnarray}
This system of equations does not yield a solution consistent with the constraints coming from the hierarchy of scales, \textit{i.e.} $u\geqslant a\geqslant b\geqslant c \geqslant d\geqslant e $. This can be seen easily by solving these equations for $u$ and $a$, which yields
\begin{eqnarray}
u= 34+0.015 (e-d)\;, \qquad a =183-b-c+1.3d-0.3e\;.
\end{eqnarray}
The maximum possible value for $u$ is obtained with $e_{max}$ and $d_{min}$, \textit{i.e.} $e=d$, and the minimum value for $a$  is obtained with, (in addition to $e=d=z\equiv\ln M_Z/\mbox{GeV}$), $b_{max}=c_{max}=a$ as
\begin{equation}
u_{max}=34\;,\qquad a_{min}=62\;.
\end{equation} 
Therefore, the system violates the required condition $u\geqslant a$, and hence does not yield a meaningful solution.

Following the same procedure, it is straightforward to show that the same situation occurs for the other configurations. Especially, the configurations $2^{\prime}-12^{\prime}$ are very similar to configuration $1^{\prime}$, which we study above, with only minor differences. For instance, the equations in case $2^{\prime}$  become
\begin{eqnarray}
u= 34+0.015 (d-e)\;, \qquad a =183-b-c+1.3e-0.3d\;.
\end{eqnarray}
In this case, in order to find $u_{max}$ we set the maximum value for $d$ and the minimum value for $e$ as $d_{max}=u$ and $e_{min}=z$. In order to find $a_{min}$ we set $b_{max}=c_{max}=a$, $e_{min}=z$, and $d_{max}=a$. Finally, we obtain
\begin{equation}
u_{max}=34.5\;,\qquad a_{min}=57\;,
\end{equation}
which clearly violates the necessary condition, $u\geqslant a$.

\subsubsection{Case 2: Configuration $13^{\prime}$ ($S\Delta \Delta \Delta  \Omega$)}
In the second example we display the calculations step by step, the mass ordering is given as 
\begin{equation}
M_U\geqslant  M_{S} \geqslant M_{\Delta_u}\geqslant M_{\Delta_d}\geqslant M_{\Delta_l} \geqslant M_{\Omega} > M_{Z}\;.
\end{equation}

\begin{table}[t]
\begin{center}
{\begin{tabular}{c|l|l}
\hline
$\vphantom{\Big|}$Interval & Active (extra) scalar dofs & RG coefficients  $\bigl[a_{1},a_{2},a_{3}\bigr]$
\\
\hline
$\vphantom{\Bigg|}$ I $(M_U-M_1)$
& $S\Delta\Delta \Delta \Omega$
& $\left[\dfrac{77}{9},-\dfrac{8}{3},-\dfrac{37}{6}\right]$
\\
\hline
$\vphantom{\Bigg|}$ II $(M_1-M_2)$
&$\Delta \Delta \Delta \Omega$
& $\left[\dfrac{17}{2},-\dfrac{19}{6},-\dfrac{13}{2}\right]$
\\
\hline
$\vphantom{\Bigg|}$ III $(M_2-M_3)$
& $\Delta \Delta \Omega$
& $\left[\dfrac{145}{18},-\dfrac{19}{6},-\dfrac{20}{3}\right]$
\\
\hline
$\vphantom{\Bigg|}$ IV $(M_3-M_4)$
& $ \Delta \Omega$
& $\left[\dfrac{137}{18},-\dfrac{19}{6},-\dfrac{41}{6}\right]$
\\
\hline
$\vphantom{\Bigg|}$ V $(M_4-M_5)$
& $ \Omega$ 
& $\left[
\dfrac{43}{6},-\dfrac{19}{6},-7
\right]$
\\
\hline
$\vphantom{\Biggl|}$   VI $(M_5-M_Z)$
& $$
& $\left[
  \dfrac{41}{6},-\dfrac{19}{6},-7
  \right]$
\\
\hline
\end{tabular}}
\caption{\label{config13}The distribution of the new scalars among the energy intervals and the corresponding RG coefficients for configuration $13^{\prime}$ ($S\Delta\Delta \Delta \Omega$), \textit{i.e.} $M_U\geqslant  M_{S} \geqslant M_{\Delta_u}\geqslant M_{\Delta_d}\geqslant M_{\Delta_l} \geqslant M_{\Omega} > M_{Z}$.}
\end{center}
\end{table}
Using Eq.~(\ref{equation1}) with the RG coefficients given in Table~\ref{config13}, we obtain
\begin{eqnarray}
\label{eqconfig13}
3265 &=& 117 u-4(b+c+d)+7a-3e\;,\nonumber\\
2289 &=& 67 u +a-e\;.
\end{eqnarray}
In order to see the inconsistency in this system, let's look at the second equation in Eq.~(\ref{eqconfig13}) solving for $u$ and $c$, which yields
\begin{eqnarray}
u=34+0.015(e-a)\;,\qquad c=183-b-d+1.3a-0.3e
\end{eqnarray}
The maximum possible value for $u$ can be found by setting $e=a$. The condition for the minimum possible value for $c$ can be easily seen by putting the second equation in the following form.
\begin{equation}
c=183+(a-b)+0.3 a-0.3e-d\;.
\end{equation}
In order to get minimum contribution from $(a-b)$ we set $a=b$ (since $a\geqslant b$). The minimum contribution for the rest of the right-hand side can be obtained for $a=a_{min}=c$ and $e=e_{max}=d=d_{max}=c$. Hence, we obtain
\begin{equation}
u_{max}=34\;,\qquad c_{min}=92\;,
\end{equation} 
which violates the necessary condition $u\geqslant c$ and hence the system does not yield an acceptable solution.

The configurations $14^{\prime}-18^{\prime}$ are very similar to the configuration $13^{\prime}$ and can be studied in the same way to find that in none of them the solutions lead to a consistent picture. The results can be read in Table~\ref{Configurations-general}.

\subsubsection{Case 3: Configuration $19^{\prime}$ ($S \Delta \Omega \Delta \Delta$)}
In the final example we display, the mass ordering is given as 
\begin{equation}
M_U\geqslant M_{S} \geqslant M_{\Delta_u} \geqslant  M_{\Omega} \geqslant M_{\Delta_d}\geqslant M_{\Delta_l}> M_{Z}\;.
\end{equation}
Using Eq.~(\ref{equation1}) with the RG coefficients given in Table~\ref{config19}, we obtain
\begin{eqnarray}
3265 &=& 117 u-4(b+d+e)+7a-3c\;,\nonumber\\
2289 &=& 67 u +a-c\;.
\end{eqnarray}
%

\begin{table}[t]
\begin{center}
{\begin{tabular}{c|l|l}
\hline
$\vphantom{\Big|}$Interval & Active (extra) scalar dofs & RG coefficients  $\bigl[a_{1},a_{2},a_{3}\bigr]$
\\
\hline
$\vphantom{\Bigg|}$ I $(M_U-M_1)$
& $S  \Delta\Omega\Delta \Delta $
& $\left[\dfrac{77}{9},-\dfrac{8}{3},-\dfrac{37}{6}\right]$
\\
\hline
$\vphantom{\Bigg|}$ II $(M_1-M_2)$
&$\Delta\Omega\Delta \Delta $
& $\left[\dfrac{17}{2},-\dfrac{19}{6},-\dfrac{13}{2}\right]$
\\
\hline
$\vphantom{\Bigg|}$ III $(M_2-M_3)$
& $\Omega\Delta \Delta $
& $\left[\dfrac{145}{18},-\dfrac{19}{6},-\dfrac{20}{3}\right]$
\\
\hline
$\vphantom{\Bigg|}$ IV $(M_3-M_4)$
& $\Delta \Delta$
& $\left[\dfrac{139}{18},-\dfrac{19}{6},-\dfrac{20}{3}\right]$
\\
\hline
$\vphantom{\Bigg|}$ V $(M_4-M_5)$
& $  \Delta$ 
& $\left[
\dfrac{131}{18},-\dfrac{19}{6},-\dfrac{41}{6}
\right]$
\\
\hline
$\vphantom{\Biggl|}$   VI $(M_5-M_Z)$
& $$
& $\left[
  \dfrac{41}{6},-\dfrac{19}{6},-7
  \right]$
\\
\hline
\end{tabular}}
\caption{\label{config19}The distribution of the new scalars among the energy intervals and the corresponding RG coefficients for configuration $19^{\prime}$ ($S\Delta \Omega \Delta\Delta $), \textit{i.e.} $M_U\geqslant M_{S} \geqslant M_{\Delta_u} \geqslant  M_{\Omega} \geqslant M_{\Delta_d}\geqslant M_{\Delta_l}> M_{Z}$.}
\end{center}
\end{table}
In order to observe the inconsistency in this system, let's solve these equations for $u$ and $b$ to find
\begin{equation}
u=34+0.015(c-a)\;,\qquad b=183-d-e+1.3a-0.3c\;.
\end{equation}
The maximum possible value for $u$ is obtained for $a=c$, whereas the minimum possible value for $b$ is achieved for $a=b=c=d=e$, leading to
\begin{equation}
u_{max}=34\;,\qquad b_{min}=92\;,
\end{equation} 
which violates the necessary condition $u\geqslant b$, hence the system does not yield a consistent solution.  Configuration $20^{\prime}$, given in Table~\ref{Configurations-general}, can be examined in a similar manner, resulting in the same situation in which there is no acceptable solution to the corresponding system of equations.

Therefore, we conclude that out of these 20 possible mass orderings there is not a single case where the gauge coupling unification can be realized at the one-loop order. Higher loop effects are expected to be suppressed and are unlikely to change this outcome.

\section{Discussions and conclusion}\label{section5}

The spectral action construction in the noncommutative geometry (NCG) framework reconciles gravity and the SM in a geometric setting, putting them on similar footings, which could possibly be considered a step towards quantum gravity. The robustness of GR and the SM can be understood from this geometric perspective.~Additionally, the NCG formalism, due to its geometric nature, might open new (perhaps non-Wilsonian) possibilities, such as decoupling of degrees of freedom and UV/IR mixing, for understanding the curious issues in the SM, such as the question of naturalness and the hierarchy problem, which has made the high energy physics community anticipate new physics at the TeV scale, likely to be on false grounds.     

One obvious question would be whether there is a fully quantized UV completion to the spectral action formalism. Without knowledge of such a completion, a spectral action can be interpreted as a classical structure, emergent from an underlying noncommutative geometry. Once that is established, the usual quantum field theory methods could be employed for the quantization and RG running for energy scales below the scale of emergence, motivated by the fact that these methods work quite well to describe Nature at low energies, accessible to the current colliders. This is indeed the approach we adopt in this paper. On the other hand, the geometric nature of this set-up brings up the question of non-local effects, which is yet to be investigated. 
  
While there are appealing features of the spectral action formalism in the NCG framework, there are also various issues within the minimal construction. The most important of these is the requirement of gauge coupling unification, which cannot be achieved by the particle content of the SM in the canonical renormalization group running. This issue is not really an indicative of a problem with the idea of the spectral action itself, but with its minimal version. After all, the spectral action formalism is not a model; it is a geometric framework that provides a toolbox for building models.~Therefore, an extension to the minimal model construction is required, in similarity to BSM physics but, in this case, in the NCG framework. Indeed, a recently proposed extension to the basic formalism accommodates (three versions of) a Pati-Salam-type model~\cite{Chamseddine:2013rta}, which is investigated from phenomenological perspective in Refs.~\cite{Aydemir:2015nfa,Aydemir:2016xtj,Aydemir:2018cbb}.
 
In a recent paper~\cite{Kurkov:2017wmx}, which is primarily based on the analysis of Ref.~\cite{DAndrea:2014ics}, it is argued that incorporating Clifford structure into the finite part of the spectral triple in the basic framework gives rise to five extra scalars. In this paper, we investigate whether these scalars can help satisfy the unification condition in this \textit{modified} minimal formalism. We study the one-loop renormalization group running in the most general case in which the extra scalars are allowed to acquire random order of masses between the electroweak scale and the presumed unification scale. We show that out of twenty possible configurations in total, depending on mass hierarchy of these additional scalars, there does \textit{not} exist even a single case that can lead to unification of the gauge couplings, as displayed in Table~\ref{Configurations-general}. Higher order loop effects are expected to be suppressed further and are unlikely to change this outcome, so are the higher order contributions in the spectral action~\cite{Devastato:2014kba}. 

In conclusion, the issue of unification in the minimal spectral action formalism is not remedied in this slightly modified scheme. Therefore, a model construction based on the spectral action principle is required to extend beyond the (modified) minimal framework with a resulting gauge sector possibly different than the one in the SM, an example of which is proposed in Ref.~\cite{Chamseddine:2013rta}, as previously mentioned. Evidently, this outcome is valid provided that the standard perturbative quantum field theory methods are appropriate to be employed all the way up to the scale of emergence of the corresponding spectral action. In case of a possible UV-complete version of the spectral action formalism, it would be conceivable to anticipate departures from the canonical RG running at scales close to the scale of emergence, which could possibly yield a self-consistent picture even with the SM field content.

\vspace{0.1cm}
\section*{Acknowledgements} 
\vspace{-0.1cm}
This work is supported by the National Natural Science Foundation of China (NSFC) under Grant No.~11505067. Author thanks Djordje Minic, Walter van Suijlekom, and Tatsu Takeuchi for their comments on the manuscript.
\vspace{-0.3cm}

\raggedright  
\bibliography{CliffordNCG}{}

\begin{thebibliography}{47}%
\makeatletter
\providecommand \@ifxundefined [1]{%
 \@ifx{#1\undefined}
}%
\providecommand \@ifnum [1]{%
 \ifnum #1\expandafter \@firstoftwo
 \else \expandafter \@secondoftwo
 \fi
}%
\providecommand \@ifx [1]{%
 \ifx #1\expandafter \@firstoftwo
 \else \expandafter \@secondoftwo
 \fi
}%
\providecommand \natexlab [1]{#1}%
\providecommand \enquote  [1]{``#1''}%
\providecommand \bibnamefont  [1]{#1}%
\providecommand \bibfnamefont [1]{#1}%
\providecommand \citenamefont [1]{#1}%
\providecommand \href@noop [0]{\@secondoftwo}%
\providecommand \href [0]{\begingroup \@sanitize@url \@href}%
\providecommand \@href[1]{\@@startlink{#1}\@@href}%
\providecommand \@@href[1]{\endgroup#1\@@endlink}%
\providecommand \@sanitize@url [0]{\catcode `\\12\catcode `\$12\catcode
  `\&12\catcode `\#12\catcode `\^12\catcode `\_12\catcode `\%12\relax}%
\providecommand \@@startlink[1]{}%
\providecommand \@@endlink[0]{}%
\providecommand \url  [0]{\begingroup\@sanitize@url \@url }%
\providecommand \@url [1]{\endgroup\@href {#1}{\urlprefix }}%
\providecommand \urlprefix  [0]{URL }%
\providecommand \Eprint [0]{\href }%
\providecommand \doibase [0]{http://dx.doi.org/}%
\providecommand \selectlanguage [0]{\@gobble}%
\providecommand \bibinfo  [0]{\@secondoftwo}%
\providecommand \bibfield  [0]{\@secondoftwo}%
\providecommand \translation [1]{[#1]}%
\providecommand \BibitemOpen [0]{}%
\providecommand \bibitemStop [0]{}%
\providecommand \bibitemNoStop [0]{.\EOS\space}%
\providecommand \EOS [0]{\spacefactor3000\relax}%
\providecommand \BibitemShut  [1]{\csname bibitem#1\endcsname}%
\let\auto@bib@innerbib\@empty
\bibitem [{\citenamefont {Connes}(1994)}]{Connes:1994yd}%
  \BibitemOpen
  \bibfield  {author} {\bibinfo {author} {\bibfnamefont {A.}~\bibnamefont
  {Connes}},\ }\href {http://www.alainconnes.org/docs/book94bigpdf.pdf} {\emph
  {\bibinfo {title} {{Noncommutative geometry}}}}\ (\bibinfo {year}
  {1994})\BibitemShut {NoStop}%
\bibitem [{\citenamefont {Connes}(1995)}]{Connes:1995tu}%
  \BibitemOpen
  \bibfield  {author} {\bibinfo {author} {\bibfnamefont {A.}~\bibnamefont
  {Connes}},\ }\href {\doibase 10.1063/1.531241} {\bibfield  {journal}
  {\bibinfo  {journal} {J. Math. Phys.}\ }\textbf {\bibinfo {volume} {36}},\
  \bibinfo {pages} {6194} (\bibinfo {year} {1995})}\BibitemShut {NoStop}%
\bibitem [{\citenamefont {Connes}(1996)}]{Connes:1996gi}%
  \BibitemOpen
  \bibfield  {author} {\bibinfo {author} {\bibfnamefont {A.}~\bibnamefont
  {Connes}},\ }\href {\doibase 10.1007/BF02506388} {\bibfield  {journal}
  {\bibinfo  {journal} {Commun. Math. Phys.}\ }\textbf {\bibinfo {volume}
  {182}},\ \bibinfo {pages} {155} (\bibinfo {year} {1996})},\ \Eprint
  {http://arxiv.org/abs/hep-th/9603053} {arXiv:hep-th/9603053 [hep-th]}
  \BibitemShut {NoStop}%
\bibitem [{\citenamefont {Connes}\ and\ \citenamefont
  {Lott}(1991)}]{Connes:1990qp}%
  \BibitemOpen
  \bibfield  {author} {\bibinfo {author} {\bibfnamefont {A.}~\bibnamefont
  {Connes}}\ and\ \bibinfo {author} {\bibfnamefont {J.}~\bibnamefont {Lott}},\
  }\bibfield  {booktitle} {\emph {\bibinfo {booktitle} {{4th Annecy Meeting on
  Theoretical Physics: Recent Advances in Field Theory Annecy, France, March
  5-9, 1990}}},\ }\href {\doibase 10.1016/0920-5632(91)90120-4} {\bibfield
  {journal} {\bibinfo  {journal} {Nucl. Phys. Proc. Suppl.}\ }\textbf {\bibinfo
  {volume} {18B}},\ \bibinfo {pages} {29} (\bibinfo {year} {1991})}\BibitemShut
  {NoStop}%
\bibitem [{\citenamefont {van Suijlekom}(2015)}]{vanSuijlekom:2015iaa}%
  \BibitemOpen
  \bibfield  {author} {\bibinfo {author} {\bibfnamefont {W.~D.}\ \bibnamefont
  {van Suijlekom}},\ }\href {\doibase 10.1007/978-94-017-9162-5} {\emph
  {\bibinfo {title} {{Noncommutative geometry and particle physics}}}},\
  Mathematical Physics Studies\ (\bibinfo  {publisher} {Springer},\ \bibinfo
  {address} {Dordrecht},\ \bibinfo {year} {2015})\BibitemShut {NoStop}%
\bibitem [{\citenamefont {Gracia-Bondia}\ \emph {et~al.}(2001)\citenamefont
  {Gracia-Bondia}, \citenamefont {Varilly},\ and\ \citenamefont
  {Figueroa}}]{GraciaBondia:2001tr}%
  \BibitemOpen
  \bibfield  {author} {\bibinfo {author} {\bibfnamefont {J.~M.}\ \bibnamefont
  {Gracia-Bondia}}, \bibinfo {author} {\bibfnamefont {J.~C.}\ \bibnamefont
  {Varilly}}, \ and\ \bibinfo {author} {\bibfnamefont {H.}~\bibnamefont
  {Figueroa}},\ }\href@noop {} {\emph {\bibinfo {title} {{Elements of
  noncommutative geometry}}}}\ (\bibinfo {year} {2001})\BibitemShut {NoStop}%
\bibitem [{\citenamefont {Chamseddine}\ and\ \citenamefont
  {Connes}(1996)}]{Chamseddine:1996rw}%
  \BibitemOpen
  \bibfield  {author} {\bibinfo {author} {\bibfnamefont {A.~H.}\ \bibnamefont
  {Chamseddine}}\ and\ \bibinfo {author} {\bibfnamefont {A.}~\bibnamefont
  {Connes}},\ }\href {\doibase 10.1103/PhysRevLett.77.4868} {\bibfield
  {journal} {\bibinfo  {journal} {Phys. Rev. Lett.}\ }\textbf {\bibinfo
  {volume} {77}},\ \bibinfo {pages} {4868} (\bibinfo {year} {1996})},\ \Eprint
  {http://arxiv.org/abs/hep-th/9606056} {arXiv:hep-th/9606056 [hep-th]}
  \BibitemShut {NoStop}%
\bibitem [{\citenamefont {Chamseddine}\ and\ \citenamefont
  {Connes}(1997)}]{Chamseddine:1996zu}%
  \BibitemOpen
  \bibfield  {author} {\bibinfo {author} {\bibfnamefont {A.~H.}\ \bibnamefont
  {Chamseddine}}\ and\ \bibinfo {author} {\bibfnamefont {A.}~\bibnamefont
  {Connes}},\ }\href {\doibase 10.1007/s002200050126} {\bibfield  {journal}
  {\bibinfo  {journal} {Commun. Math. Phys.}\ }\textbf {\bibinfo {volume}
  {186}},\ \bibinfo {pages} {731} (\bibinfo {year} {1997})},\ \Eprint
  {http://arxiv.org/abs/hep-th/9606001} {arXiv:hep-th/9606001 [hep-th]}
  \BibitemShut {NoStop}%
\bibitem [{\citenamefont {Chamseddine}\ \emph {et~al.}(2007)\citenamefont
  {Chamseddine}, \citenamefont {Connes},\ and\ \citenamefont
  {Marcolli}}]{Chamseddine:2006ep}%
  \BibitemOpen
  \bibfield  {author} {\bibinfo {author} {\bibfnamefont {A.~H.}\ \bibnamefont
  {Chamseddine}}, \bibinfo {author} {\bibfnamefont {A.}~\bibnamefont {Connes}},
  \ and\ \bibinfo {author} {\bibfnamefont {M.}~\bibnamefont {Marcolli}},\
  }\href {\doibase 10.4310/ATMP.2007.v11.n6.a3} {\bibfield  {journal} {\bibinfo
   {journal} {Adv. Theor. Math. Phys.}\ }\textbf {\bibinfo {volume} {11}},\
  \bibinfo {pages} {991} (\bibinfo {year} {2007})},\ \Eprint
  {http://arxiv.org/abs/hep-th/0610241} {arXiv:hep-th/0610241 [hep-th]}
  \BibitemShut {NoStop}%
\bibitem [{\citenamefont {Chamseddine}\ and\ \citenamefont
  {Connes}(2008)}]{Chamseddine:2007hz}%
  \BibitemOpen
  \bibfield  {author} {\bibinfo {author} {\bibfnamefont {A.~H.}\ \bibnamefont
  {Chamseddine}}\ and\ \bibinfo {author} {\bibfnamefont {A.}~\bibnamefont
  {Connes}},\ }\href {\doibase 10.1016/j.geomphys.2007.09.011} {\bibfield
  {journal} {\bibinfo  {journal} {J. Geom. Phys.}\ }\textbf {\bibinfo {volume}
  {58}},\ \bibinfo {pages} {38} (\bibinfo {year} {2008})},\ \Eprint
  {http://arxiv.org/abs/0706.3688} {arXiv:0706.3688 [hep-th]} \BibitemShut
  {NoStop}%
\bibitem [{\citenamefont {Chamseddine}\ and\ \citenamefont
  {Connes}(2010)}]{Chamseddine:2010ud}%
  \BibitemOpen
  \bibfield  {author} {\bibinfo {author} {\bibfnamefont {A.~H.}\ \bibnamefont
  {Chamseddine}}\ and\ \bibinfo {author} {\bibfnamefont {A.}~\bibnamefont
  {Connes}},\ }\href {\doibase 10.1002/prop.201000069} {\bibfield  {journal}
  {\bibinfo  {journal} {Fortsch. Phys.}\ }\textbf {\bibinfo {volume} {58}},\
  \bibinfo {pages} {553} (\bibinfo {year} {2010})},\ \Eprint
  {http://arxiv.org/abs/1004.0464} {arXiv:1004.0464 [hep-th]} \BibitemShut
  {NoStop}%
\bibitem [{\citenamefont {Chamseddine}\ and\ \citenamefont
  {Connes}(2012)}]{Chamseddine:2012sw}%
  \BibitemOpen
  \bibfield  {author} {\bibinfo {author} {\bibfnamefont {A.~H.}\ \bibnamefont
  {Chamseddine}}\ and\ \bibinfo {author} {\bibfnamefont {A.}~\bibnamefont
  {Connes}},\ }\href {\doibase 10.1007/JHEP09(2012)104} {\bibfield  {journal}
  {\bibinfo  {journal} {JHEP}\ }\textbf {\bibinfo {volume} {09}},\ \bibinfo
  {pages} {104} (\bibinfo {year} {2012})},\ \Eprint
  {http://arxiv.org/abs/1208.1030} {arXiv:1208.1030 [hep-ph]} \BibitemShut
  {NoStop}%
\bibitem [{\citenamefont {Connes}\ and\ \citenamefont
  {Marcolli}(2007)}]{Connes:2007book}%
  \BibitemOpen
  \bibfield  {author} {\bibinfo {author} {\bibfnamefont {A.}~\bibnamefont
  {Connes}}\ and\ \bibinfo {author} {\bibfnamefont {M.}~\bibnamefont
  {Marcolli}},\ }\href {http://www.alainconnes.org/docs/bookwebfinal.pdf}
  {\emph {\bibinfo {title} {{Noncommutative Geometry, Quantum Fields and
  Motives}}}},\ 1414300\ (\bibinfo  {publisher} {American Mathematical
  Society},\ \bibinfo {year} {2007})\BibitemShut {NoStop}%
\bibitem [{\citenamefont {Marcolli}(2018)}]{Marcolli:2018uea}%
  \BibitemOpen
  \bibfield  {author} {\bibinfo {author} {\bibfnamefont {M.}~\bibnamefont
  {Marcolli}},\ }\href {\doibase 10.1142/10335} {\emph {\bibinfo {title}
  {{Noncommutative Cosmology}}}}\ (\bibinfo  {publisher} {WSP},\ \bibinfo
  {year} {2018})\BibitemShut {NoStop}%
\bibitem [{\citenamefont {Lizzi}(2018)}]{Lizzi:2018dah}%
  \BibitemOpen
  \bibfield  {author} {\bibinfo {author} {\bibfnamefont {F.}~\bibnamefont
  {Lizzi}},\ }\bibfield  {booktitle} {\emph {\bibinfo {booktitle}
  {{Proceedings, 17th Hellenic School and Workshops on Elementary Particle
  Physics and Gravity (CORFU2017): Corfu, Greece, September 2-28, 2017}}},\
  }\href {\doibase 10.22323/1.318.0133} {\bibfield  {journal} {\bibinfo
  {journal} {PoS}\ }\textbf {\bibinfo {volume} {CORFU2017}},\ \bibinfo {pages}
  {133} (\bibinfo {year} {2018})},\ \Eprint {http://arxiv.org/abs/1805.00411}
  {arXiv:1805.00411 [hep-th]} \BibitemShut {NoStop}%
\bibitem [{\citenamefont {Chamseddine}\ \emph
  {et~al.}(2013{\natexlab{a}})\citenamefont {Chamseddine}, \citenamefont
  {Connes},\ and\ \citenamefont {van Suijlekom}}]{Chamseddine:2013rta}%
  \BibitemOpen
  \bibfield  {author} {\bibinfo {author} {\bibfnamefont {A.~H.}\ \bibnamefont
  {Chamseddine}}, \bibinfo {author} {\bibfnamefont {A.}~\bibnamefont {Connes}},
  \ and\ \bibinfo {author} {\bibfnamefont {W.~D.}\ \bibnamefont {van
  Suijlekom}},\ }\href {\doibase 10.1007/JHEP11(2013)132} {\bibfield  {journal}
  {\bibinfo  {journal} {JHEP}\ }\textbf {\bibinfo {volume} {11}},\ \bibinfo
  {pages} {132} (\bibinfo {year} {2013}{\natexlab{a}})},\ \Eprint
  {http://arxiv.org/abs/1304.8050} {arXiv:1304.8050 [hep-th]} \BibitemShut
  {NoStop}%
\bibitem [{\citenamefont {Pati}\ and\ \citenamefont
  {Salam}(1974)}]{Pati:1974yy}%
  \BibitemOpen
  \bibfield  {author} {\bibinfo {author} {\bibfnamefont {J.~C.}\ \bibnamefont
  {Pati}}\ and\ \bibinfo {author} {\bibfnamefont {A.}~\bibnamefont {Salam}},\
  }\href {\doibase 10.1103/PhysRevD.10.275, 10.1103/PhysRevD.11.703.2}
  {\bibfield  {journal} {\bibinfo  {journal} {Phys. Rev.}\ }\textbf {\bibinfo
  {volume} {D10}},\ \bibinfo {pages} {275} (\bibinfo {year} {1974})},\ \bibinfo
  {note} {[Erratum: Phys. Rev. D11, 703 (1975)]}\BibitemShut {NoStop}%
\bibitem [{\citenamefont {Mohapatra}\ and\ \citenamefont
  {Pati}(1975{\natexlab{a}})}]{Mohapatra:1974gc}%
  \BibitemOpen
  \bibfield  {author} {\bibinfo {author} {\bibfnamefont {R.~N.}\ \bibnamefont
  {Mohapatra}}\ and\ \bibinfo {author} {\bibfnamefont {J.~C.}\ \bibnamefont
  {Pati}},\ }\href {\doibase 10.1103/PhysRevD.11.2558} {\bibfield  {journal}
  {\bibinfo  {journal} {Phys. Rev.}\ }\textbf {\bibinfo {volume} {D11}},\
  \bibinfo {pages} {2558} (\bibinfo {year} {1975}{\natexlab{a}})}\BibitemShut
  {NoStop}%
\bibitem [{\citenamefont {Mohapatra}\ and\ \citenamefont
  {Pati}(1975{\natexlab{b}})}]{Mohapatra:1974hk}%
  \BibitemOpen
  \bibfield  {author} {\bibinfo {author} {\bibfnamefont {R.~N.}\ \bibnamefont
  {Mohapatra}}\ and\ \bibinfo {author} {\bibfnamefont {J.~C.}\ \bibnamefont
  {Pati}},\ }\href {\doibase 10.1103/PhysRevD.11.566} {\bibfield  {journal}
  {\bibinfo  {journal} {Phys. Rev.}\ }\textbf {\bibinfo {volume} {D11}},\
  \bibinfo {pages} {566} (\bibinfo {year} {1975}{\natexlab{b}})}\BibitemShut
  {NoStop}%
\bibitem [{\citenamefont {Senjanovic}\ and\ \citenamefont
  {Mohapatra}(1975)}]{Senjanovic:1975rk}%
  \BibitemOpen
  \bibfield  {author} {\bibinfo {author} {\bibfnamefont {G.}~\bibnamefont
  {Senjanovic}}\ and\ \bibinfo {author} {\bibfnamefont {R.~N.}\ \bibnamefont
  {Mohapatra}},\ }\href {\doibase 10.1103/PhysRevD.12.1502} {\bibfield
  {journal} {\bibinfo  {journal} {Phys. Rev.}\ }\textbf {\bibinfo {volume}
  {D12}},\ \bibinfo {pages} {1502} (\bibinfo {year} {1975})}\BibitemShut
  {NoStop}%
\bibitem [{\citenamefont {Chamseddine}\ \emph {et~al.}(2015)\citenamefont
  {Chamseddine}, \citenamefont {Connes},\ and\ \citenamefont {van
  Suijlekom}}]{Chamseddine:2015ata}%
  \BibitemOpen
  \bibfield  {author} {\bibinfo {author} {\bibfnamefont {A.~H.}\ \bibnamefont
  {Chamseddine}}, \bibinfo {author} {\bibfnamefont {A.}~\bibnamefont {Connes}},
  \ and\ \bibinfo {author} {\bibfnamefont {W.~D.}\ \bibnamefont {van
  Suijlekom}},\ }\href {\doibase 10.1007/JHEP11(2015)011} {\bibfield  {journal}
  {\bibinfo  {journal} {JHEP}\ }\textbf {\bibinfo {volume} {11}},\ \bibinfo
  {pages} {011} (\bibinfo {year} {2015})},\ \Eprint
  {http://arxiv.org/abs/1507.08161} {arXiv:1507.08161 [hep-ph]} \BibitemShut
  {NoStop}%
\bibitem [{\citenamefont {Aydemir}\ \emph
  {et~al.}(2016{\natexlab{a}})\citenamefont {Aydemir}, \citenamefont {Minic},
  \citenamefont {Sun},\ and\ \citenamefont {Takeuchi}}]{Aydemir:2015nfa}%
  \BibitemOpen
  \bibfield  {author} {\bibinfo {author} {\bibfnamefont {U.}~\bibnamefont
  {Aydemir}}, \bibinfo {author} {\bibfnamefont {D.}~\bibnamefont {Minic}},
  \bibinfo {author} {\bibfnamefont {C.}~\bibnamefont {Sun}}, \ and\ \bibinfo
  {author} {\bibfnamefont {T.}~\bibnamefont {Takeuchi}},\ }\href {\doibase
  10.1142/S0217751X15502231} {\bibfield  {journal} {\bibinfo  {journal} {Int.
  J. Mod. Phys.}\ }\textbf {\bibinfo {volume} {A31}},\ \bibinfo {pages}
  {1550223} (\bibinfo {year} {2016}{\natexlab{a}})},\ \Eprint
  {http://arxiv.org/abs/1509.01606} {ArXiv:1509.01606 [hep-ph]} \BibitemShut
  {NoStop}%
\bibitem [{\citenamefont {Aydemir}\ \emph
  {et~al.}(2016{\natexlab{b}})\citenamefont {Aydemir}, \citenamefont {Minic},
  \citenamefont {Sun},\ and\ \citenamefont {Takeuchi}}]{Aydemir:2016xtj}%
  \BibitemOpen
  \bibfield  {author} {\bibinfo {author} {\bibfnamefont {U.}~\bibnamefont
  {Aydemir}}, \bibinfo {author} {\bibfnamefont {D.}~\bibnamefont {Minic}},
  \bibinfo {author} {\bibfnamefont {C.}~\bibnamefont {Sun}}, \ and\ \bibinfo
  {author} {\bibfnamefont {T.}~\bibnamefont {Takeuchi}},\ }\href {\doibase
  10.1142/S0217732316501017} {\bibfield  {journal} {\bibinfo  {journal} {Mod.
  Phys. Lett.}\ }\textbf {\bibinfo {volume} {A31}},\ \bibinfo {pages} {1650101}
  (\bibinfo {year} {2016}{\natexlab{b}})},\ \Eprint
  {http://arxiv.org/abs/1603.01756} {ArXiv:1603.01756 [hep-ph]} \BibitemShut
  {NoStop}%
\bibitem [{\citenamefont {Aydemir}\ \emph {et~al.}(2018)\citenamefont
  {Aydemir}, \citenamefont {Minic}, \citenamefont {Sun},\ and\ \citenamefont
  {Takeuchi}}]{Aydemir:2018cbb}%
  \BibitemOpen
  \bibfield  {author} {\bibinfo {author} {\bibfnamefont {U.}~\bibnamefont
  {Aydemir}}, \bibinfo {author} {\bibfnamefont {D.}~\bibnamefont {Minic}},
  \bibinfo {author} {\bibfnamefont {C.}~\bibnamefont {Sun}}, \ and\ \bibinfo
  {author} {\bibfnamefont {T.}~\bibnamefont {Takeuchi}},\ }\href {\doibase
  10.1007/JHEP09(2018)117} {\bibfield  {journal} {\bibinfo  {journal} {JHEP}\
  }\textbf {\bibinfo {volume} {09}},\ \bibinfo {pages} {117} (\bibinfo {year}
  {2018})},\ \Eprint {http://arxiv.org/abs/1804.05844} {ArXiv:1804.05844
  [hep-ph]} \BibitemShut {NoStop}%
\bibitem [{\citenamefont {Kurkov}\ and\ \citenamefont
  {Lizzi}(2018)}]{Kurkov:2017wmx}%
  \BibitemOpen
  \bibfield  {author} {\bibinfo {author} {\bibfnamefont {M.~A.}\ \bibnamefont
  {Kurkov}}\ and\ \bibinfo {author} {\bibfnamefont {F.}~\bibnamefont {Lizzi}},\
  }\href {\doibase 10.1103/PhysRevD.97.085024} {\bibfield  {journal} {\bibinfo
  {journal} {Phys. Rev.}\ }\textbf {\bibinfo {volume} {D97}},\ \bibinfo {pages}
  {085024} (\bibinfo {year} {2018})},\ \Eprint
  {http://arxiv.org/abs/1801.00260} {arXiv:1801.00260 [hep-th]} \BibitemShut
  {NoStop}%
\bibitem [{\citenamefont {D'Andrea}\ and\ \citenamefont
  {Dabrowski}(2014)}]{DAndrea:2014ics}%
  \BibitemOpen
  \bibfield  {author} {\bibinfo {author} {\bibfnamefont {F.}~\bibnamefont
  {D'Andrea}}\ and\ \bibinfo {author} {\bibfnamefont {L.}~\bibnamefont
  {Dabrowski}},\ }\href@noop {} {\  (\bibinfo {year} {2014})},\ \Eprint
  {http://arxiv.org/abs/1501.00156} {arXiv:1501.00156 [math-ph]} \BibitemShut
  {NoStop}%
\bibitem [{\citenamefont {Aaij}\ \emph {et~al.}(2014)\citenamefont {Aaij} \emph
  {et~al.}}]{Aaij:2014ora}%
  \BibitemOpen
  \bibfield  {author} {\bibinfo {author} {\bibfnamefont {R.}~\bibnamefont
  {Aaij}} \emph {et~al.} (\bibinfo {collaboration} {LHCb}),\ }\href {\doibase
  10.1103/PhysRevLett.113.151601} {\bibfield  {journal} {\bibinfo  {journal}
  {Phys. Rev. Lett.}\ }\textbf {\bibinfo {volume} {113}},\ \bibinfo {pages}
  {151601} (\bibinfo {year} {2014})},\ \Eprint {http://arxiv.org/abs/1406.6482}
  {arXiv:1406.6482 [hep-ex]} \BibitemShut {NoStop}%
\bibitem [{\citenamefont {Aaij}\ \emph {et~al.}(2017)\citenamefont {Aaij} \emph
  {et~al.}}]{Aaij:2017vbb}%
  \BibitemOpen
  \bibfield  {author} {\bibinfo {author} {\bibfnamefont {R.}~\bibnamefont
  {Aaij}} \emph {et~al.} (\bibinfo {collaboration} {LHCb}),\ }\href {\doibase
  10.1007/JHEP08(2017)055} {\bibfield  {journal} {\bibinfo  {journal} {JHEP}\
  }\textbf {\bibinfo {volume} {08}},\ \bibinfo {pages} {055} (\bibinfo {year}
  {2017})},\ \Eprint {http://arxiv.org/abs/1705.05802} {arXiv:1705.05802
  [hep-ex]} \BibitemShut {NoStop}%
\bibitem [{\citenamefont {Aaij}\ \emph
  {et~al.}(2018{\natexlab{a}})\citenamefont {Aaij} \emph
  {et~al.}}]{Aaij:2017uff}%
  \BibitemOpen
  \bibfield  {author} {\bibinfo {author} {\bibfnamefont {R.}~\bibnamefont
  {Aaij}} \emph {et~al.} (\bibinfo {collaboration} {LHCb}),\ }\href {\doibase
  10.1103/PhysRevLett.120.171802} {\bibfield  {journal} {\bibinfo  {journal}
  {Phys. Rev. Lett.}\ }\textbf {\bibinfo {volume} {120}},\ \bibinfo {pages}
  {171802} (\bibinfo {year} {2018}{\natexlab{a}})},\ \Eprint
  {http://arxiv.org/abs/1708.08856} {arXiv:1708.08856 [hep-ex]} \BibitemShut
  {NoStop}%
\bibitem [{\citenamefont {Aaij}\ \emph {et~al.}(2015)\citenamefont {Aaij} \emph
  {et~al.}}]{Aaij:2015yra}%
  \BibitemOpen
  \bibfield  {author} {\bibinfo {author} {\bibfnamefont {R.}~\bibnamefont
  {Aaij}} \emph {et~al.} (\bibinfo {collaboration} {{LHCb}}),\ }\href {\doibase
  10.1103/PhysRevLett.115.159901, 10.1103/PhysRevLett.115.111803} {\bibfield
  {journal} {\bibinfo  {journal} {Phys. Rev. Lett.}\ }\textbf {\bibinfo
  {volume} {115}},\ \bibinfo {pages} {111803} (\bibinfo {year} {2015})},\
  \bibinfo {note} {[Erratum: Phys. Rev. Lett. 115, no.15, 159901 (2015)]},\
  \Eprint {http://arxiv.org/abs/1506.08614} {arXiv:1506.08614 [hep-ex]}
  \BibitemShut {NoStop}%
\bibitem [{\citenamefont {Aaij}\ \emph
  {et~al.}(2018{\natexlab{b}})\citenamefont {Aaij} \emph
  {et~al.}}]{Aaij:2017deq}%
  \BibitemOpen
  \bibfield  {author} {\bibinfo {author} {\bibfnamefont {R.}~\bibnamefont
  {Aaij}} \emph {et~al.} (\bibinfo {collaboration} {LHCb}),\ }\href {\doibase
  10.1103/PhysRevD.97.072013} {\bibfield  {journal} {\bibinfo  {journal} {Phys.
  Rev.}\ }\textbf {\bibinfo {volume} {D97}},\ \bibinfo {pages} {072013}
  (\bibinfo {year} {2018}{\natexlab{b}})},\ \Eprint
  {http://arxiv.org/abs/1711.02505} {arXiv:1711.02505 [hep-ex]} \BibitemShut
  {NoStop}%
\bibitem [{\citenamefont {Huschle}\ \emph {et~al.}(2015)\citenamefont {Huschle}
  \emph {et~al.}}]{Huschle:2015rga}%
  \BibitemOpen
  \bibfield  {author} {\bibinfo {author} {\bibfnamefont {M.}~\bibnamefont
  {Huschle}} \emph {et~al.} (\bibinfo {collaboration} {Belle}),\ }\href
  {\doibase 10.1103/PhysRevD.92.072014} {\bibfield  {journal} {\bibinfo
  {journal} {Phys. Rev.}\ }\textbf {\bibinfo {volume} {D92}},\ \bibinfo {pages}
  {072014} (\bibinfo {year} {2015})},\ \Eprint
  {http://arxiv.org/abs/1507.03233} {arXiv:1507.03233 [hep-ex]} \BibitemShut
  {NoStop}%
\bibitem [{\citenamefont {Sato}\ \emph {et~al.}(2016)\citenamefont {Sato} \emph
  {et~al.}}]{Sato:2016svk}%
  \BibitemOpen
  \bibfield  {author} {\bibinfo {author} {\bibfnamefont {Y.}~\bibnamefont
  {Sato}} \emph {et~al.} (\bibinfo {collaboration} {Belle}),\ }\href {\doibase
  10.1103/PhysRevD.94.072007} {\bibfield  {journal} {\bibinfo  {journal} {Phys.
  Rev.}\ }\textbf {\bibinfo {volume} {D94}},\ \bibinfo {pages} {072007}
  (\bibinfo {year} {2016})},\ \Eprint {http://arxiv.org/abs/1607.07923}
  {arXiv:1607.07923 [hep-ex]} \BibitemShut {NoStop}%
\bibitem [{\citenamefont {Hirose}\ \emph {et~al.}(2017)\citenamefont {Hirose}
  \emph {et~al.}}]{Hirose:2016wfn}%
  \BibitemOpen
  \bibfield  {author} {\bibinfo {author} {\bibfnamefont {S.}~\bibnamefont
  {Hirose}} \emph {et~al.} (\bibinfo {collaboration} {Belle}),\ }\href
  {\doibase 10.1103/PhysRevLett.118.211801} {\bibfield  {journal} {\bibinfo
  {journal} {Phys. Rev. Lett.}\ }\textbf {\bibinfo {volume} {118}},\ \bibinfo
  {pages} {211801} (\bibinfo {year} {2017})},\ \Eprint
  {http://arxiv.org/abs/1612.00529} {arXiv:1612.00529 [hep-ex]} \BibitemShut
  {NoStop}%
\bibitem [{\citenamefont {Lees}\ \emph {et~al.}(2012)\citenamefont {Lees} \emph
  {et~al.}}]{Lees:2012xj}%
  \BibitemOpen
  \bibfield  {author} {\bibinfo {author} {\bibfnamefont {J.~P.}\ \bibnamefont
  {Lees}} \emph {et~al.} (\bibinfo {collaboration} {BaBar}),\ }\href {\doibase
  10.1103/PhysRevLett.109.101802} {\bibfield  {journal} {\bibinfo  {journal}
  {Phys. Rev. Lett.}\ }\textbf {\bibinfo {volume} {109}},\ \bibinfo {pages}
  {101802} (\bibinfo {year} {2012})},\ \Eprint {http://arxiv.org/abs/1205.5442}
  {arXiv:1205.5442 [hep-ex]} \BibitemShut {NoStop}%
\bibitem [{\citenamefont {Lees}\ \emph {et~al.}(2013)\citenamefont {Lees} \emph
  {et~al.}}]{Lees:2013uzd}%
  \BibitemOpen
  \bibfield  {author} {\bibinfo {author} {\bibfnamefont {J.~P.}\ \bibnamefont
  {Lees}} \emph {et~al.} (\bibinfo {collaboration} {BaBar}),\ }\href {\doibase
  10.1103/PhysRevD.88.072012} {\bibfield  {journal} {\bibinfo  {journal} {Phys.
  Rev.}\ }\textbf {\bibinfo {volume} {D88}},\ \bibinfo {pages} {072012}
  (\bibinfo {year} {2013})},\ \Eprint {http://arxiv.org/abs/1303.0571}
  {arXiv:1303.0571 [hep-ex]} \BibitemShut {NoStop}%
\bibitem [{\citenamefont {Eckstein}\ and\ \citenamefont
  {Iochum}(2019)}]{Eckstein:2019dcb}%
  \BibitemOpen
  \bibfield  {author} {\bibinfo {author} {\bibfnamefont {M.}~\bibnamefont
  {Eckstein}}\ and\ \bibinfo {author} {\bibfnamefont {B.}~\bibnamefont
  {Iochum}},\ }\href {\doibase 10.1007/978-3-319-94788-4} {\  (\bibinfo {year}
  {2019}),\ 10.1007/978-3-319-94788-4},\ \Eprint
  {http://arxiv.org/abs/1902.05306} {arXiv:1902.05306 [math-ph]} \BibitemShut
  {NoStop}%
\bibitem [{\citenamefont {Chamseddine}\ \emph
  {et~al.}(2013{\natexlab{b}})\citenamefont {Chamseddine}, \citenamefont
  {Connes},\ and\ \citenamefont {van Suijlekom}}]{Chamseddine:2013kza}%
  \BibitemOpen
  \bibfield  {author} {\bibinfo {author} {\bibfnamefont {A.~H.}\ \bibnamefont
  {Chamseddine}}, \bibinfo {author} {\bibfnamefont {A.}~\bibnamefont {Connes}},
  \ and\ \bibinfo {author} {\bibfnamefont {W.~D.}\ \bibnamefont {van
  Suijlekom}},\ }\href {\doibase 10.1016/j.geomphys.2013.06.006} {\bibfield
  {journal} {\bibinfo  {journal} {J. Geom. Phys.}\ }\textbf {\bibinfo {volume}
  {73}},\ \bibinfo {pages} {222} (\bibinfo {year} {2013}{\natexlab{b}})},\
  \Eprint {http://arxiv.org/abs/1304.7583} {arXiv:1304.7583 [math-ph]}
  \BibitemShut {NoStop}%
\bibitem [{\citenamefont {D'Andrea}\ \emph {et~al.}(2016)\citenamefont
  {D'Andrea}, \citenamefont {Kurkov},\ and\ \citenamefont
  {Lizzi}}]{DAndrea:2016hyl}%
  \BibitemOpen
  \bibfield  {author} {\bibinfo {author} {\bibfnamefont {F.}~\bibnamefont
  {D'Andrea}}, \bibinfo {author} {\bibfnamefont {M.~A.}\ \bibnamefont
  {Kurkov}}, \ and\ \bibinfo {author} {\bibfnamefont {F.}~\bibnamefont
  {Lizzi}},\ }\href {\doibase 10.1103/PhysRevD.94.025030} {\bibfield  {journal}
  {\bibinfo  {journal} {Phys. Rev.}\ }\textbf {\bibinfo {volume} {D94}},\
  \bibinfo {pages} {025030} (\bibinfo {year} {2016})},\ \Eprint
  {http://arxiv.org/abs/1605.03231} {arXiv:1605.03231 [hep-th]} \BibitemShut
  {NoStop}%
\bibitem [{\citenamefont {Gilkey}(1994)}]{gilkey1994invariance}%
  \BibitemOpen
  \bibfield  {author} {\bibinfo {author} {\bibfnamefont {P.}~\bibnamefont
  {Gilkey}},\ }\href {https://books.google.com/books?id=hG\_vAAAAMAAJ} {\emph
  {\bibinfo {title} {Invariance Theory: The Heat Equation and the Atiyah-Singer
  Index Theorem}}},\ Studies in Advanced Mathematics\ (\bibinfo  {publisher}
  {CRC-Press},\ \bibinfo {year} {1994})\BibitemShut {NoStop}%
\bibitem [{\citenamefont {Vassilevich}(2003)}]{Vassilevich:2003xt}%
  \BibitemOpen
  \bibfield  {author} {\bibinfo {author} {\bibfnamefont {D.~V.}\ \bibnamefont
  {Vassilevich}},\ }\href {\doibase 10.1016/j.physrep.2003.09.002} {\bibfield
  {journal} {\bibinfo  {journal} {Phys. Rept.}\ }\textbf {\bibinfo {volume}
  {388}},\ \bibinfo {pages} {279} (\bibinfo {year} {2003})},\ \Eprint
  {http://arxiv.org/abs/hep-th/0306138} {arXiv:hep-th/0306138 [hep-th]}
  \BibitemShut {NoStop}%
\bibitem [{\citenamefont {Jones}(1982)}]{Jones:1981we}%
  \BibitemOpen
  \bibfield  {author} {\bibinfo {author} {\bibfnamefont {D.~R.~T.}\
  \bibnamefont {Jones}},\ }\href {\doibase 10.1103/PhysRevD.25.581} {\bibfield
  {journal} {\bibinfo  {journal} {Phys. Rev.}\ }\textbf {\bibinfo {volume}
  {D25}},\ \bibinfo {pages} {581} (\bibinfo {year} {1982})}\BibitemShut
  {NoStop}%
\bibitem [{\citenamefont {Lindner}\ and\ \citenamefont
  {Weiser}(1996)}]{Lindner:1996tf}%
  \BibitemOpen
  \bibfield  {author} {\bibinfo {author} {\bibfnamefont {M.}~\bibnamefont
  {Lindner}}\ and\ \bibinfo {author} {\bibfnamefont {M.}~\bibnamefont
  {Weiser}},\ }\href {\doibase 10.1016/0370-2693(96)00775-7} {\bibfield
  {journal} {\bibinfo  {journal} {Phys. Lett.}\ }\textbf {\bibinfo {volume}
  {B383}},\ \bibinfo {pages} {405} (\bibinfo {year} {1996})},\ \Eprint
  {http://arxiv.org/abs/hep-ph/9605353} {arXiv:hep-ph/9605353 [hep-ph]}
  \BibitemShut {NoStop}%
\bibitem [{\citenamefont {Patrignani}\ \emph {et~al.}(2016)\citenamefont
  {Patrignani} \emph {et~al.}}]{Patrignani:2016xqp}%
  \BibitemOpen
  \bibfield  {author} {\bibinfo {author} {\bibfnamefont {C.}~\bibnamefont
  {Patrignani}} \emph {et~al.} (\bibinfo {collaboration} {Particle Data
  Group}),\ }\href {\doibase 10.1088/1674-1137/40/10/100001} {\bibfield
  {journal} {\bibinfo  {journal} {Chin. Phys.}\ }\textbf {\bibinfo {volume}
  {C40}},\ \bibinfo {pages} {100001} (\bibinfo {year} {2016})}\BibitemShut
  {NoStop}%
\bibitem [{\citenamefont {Schael}\ \emph {et~al.}(2006)\citenamefont {Schael}
  \emph {et~al.}}]{ALEPH:2005ab}%
  \BibitemOpen
  \bibfield  {author} {\bibinfo {author} {\bibfnamefont {S.}~\bibnamefont
  {Schael}} \emph {et~al.} (\bibinfo {collaboration} {SLD Electroweak Group,
  DELPHI, ALEPH, SLD, SLD Heavy Flavour Group, OPAL, LEP Electroweak Working
  Group, L3}),\ }\href {\doibase 10.1016/j.physrep.2005.12.006} {\bibfield
  {journal} {\bibinfo  {journal} {Phys. Rept.}\ }\textbf {\bibinfo {volume}
  {427}},\ \bibinfo {pages} {257} (\bibinfo {year} {2006})},\ \Eprint
  {http://arxiv.org/abs/hep-ex/0509008} {arXiv:hep-ex/0509008 [hep-ex]}
  \BibitemShut {NoStop}%
\bibitem [{\citenamefont {Aydemir}\ and\ \citenamefont
  {Mandal}(2017)}]{Aydemir:2016qqj}%
  \BibitemOpen
  \bibfield  {author} {\bibinfo {author} {\bibfnamefont {U.}~\bibnamefont
  {Aydemir}}\ and\ \bibinfo {author} {\bibfnamefont {T.}~\bibnamefont
  {Mandal}},\ }\href {\doibase 10.1155/2017/7498795} {\bibfield  {journal}
  {\bibinfo  {journal} {Adv. High Energy Phys.}\ }\textbf {\bibinfo {volume}
  {2017}},\ \bibinfo {pages} {7498795} (\bibinfo {year} {2017})},\ \Eprint
  {http://arxiv.org/abs/1601.06761} {arXiv:1601.06761 [hep-ph]} \BibitemShut
  {NoStop}%
\bibitem [{\citenamefont {Devastato}\ \emph {et~al.}(2015)\citenamefont
  {Devastato}, \citenamefont {Lizzi}, \citenamefont {Flores},\ and\
  \citenamefont {Vassilevich}}]{Devastato:2014kba}%
  \BibitemOpen
  \bibfield  {author} {\bibinfo {author} {\bibfnamefont {A.}~\bibnamefont
  {Devastato}}, \bibinfo {author} {\bibfnamefont {F.}~\bibnamefont {Lizzi}},
  \bibinfo {author} {\bibfnamefont {C.~V.}\ \bibnamefont {Flores}}, \ and\
  \bibinfo {author} {\bibfnamefont {D.}~\bibnamefont {Vassilevich}},\ }\href
  {\doibase 10.1142/S0217751X15500335} {\bibfield  {journal} {\bibinfo
  {journal} {Int. J. Mod. Phys.}\ }\textbf {\bibinfo {volume} {A30}},\ \bibinfo
  {pages} {1550033} (\bibinfo {year} {2015})},\ \Eprint
  {http://arxiv.org/abs/1410.6624} {arXiv:1410.6624 [hep-ph]} \BibitemShut
  {NoStop}%
\end{thebibliography}%
\bibliographystyle{apsrev4-1}

\end{document}